\documentclass{article} 
\usepackage[left=2.5cm,right=2.5cm,top=2.5cm,bottom=2.5cm,bindingoffset=0cm]{geometry}
\usepackage{cmap}
\usepackage[english]{babel} 
\usepackage{indentfirst}
\usepackage{amssymb,amsmath}
\usepackage{graphicx}
\usepackage[hyphens]{url} 
\usepackage{amsthm}
\usepackage{pgfplots}
\usepackage{float}
\usepackage{hyperref}
\usepackage{xcolor}
\usepackage{comment}
\usepackage{algorithm}
\usepackage{algpseudocode}

\theoremstyle{definition}

\title{Feedforward Inversion Control of DC/DC Dual-Bridge Series Resonant Converter in Buck and Boost Modes}
\date{\today}
\author{Alex Borisevich, \href{mailto:akpc806b@gmail.com}{akpc806b@gmail.com}, \\ 
Filipp Gleyzer, \href{mailto:gleyzerf@smccd.edu}{gleyzerf@smccd.edu}, \\
Valeria Gavrilenko, \href{mailto:gavv00@vse.cz}{gavv00@vse.cz}
}

\DeclareMathOperator{\atan}{atan}
\DeclareMathOperator{\atantwo}{atan2}
\DeclareMathOperator{\acos}{acos}

\begin{document}

\maketitle

\section*{Abstract}

In the paper, a nonlinear inversion technique for the steady-state model of the active dual-bridge series resonant converter is presented. The obtained control strategy allows cycle averaged output current regulation and performs waveform alignment for the controllable achievement of ZVS and synchronous rectification. The control is valid both for voltage buck and boost operating modes, as well as for low-power operation at a fixed frequency. Robustness of the control is studied by simulations with external linear control loops.

\section{Introduction}

The dual-bridge (DB) series resonant converter (SRC) in comparison to many well known isolated DC/DC topologies like LLC or non-resonant dual active bridge has some unique features due to controlled secondary-side bridge and the presence of resonant LC tank, such as close to sinusoidal transformer current and a capability of bidirectional power flow and voltage boost (step-up) operation [1]. Such type of converter topology is particularly promising for electric vehicles chargers applications [1-3], including off-board charging stations and on-board charger modules with an ability for bidirectional vehicle-to-grid power flow. Additionally, many other power electronics applications like battery energy storage and DC line power transmission are target areas of DB SRC topology.

The modulation scheme for DB SRC consists of driving all the switches in the two bridges work with 50

The modulation scheme for DB SRC usually enables pure ZVS on the primary side, and combined ZVS/ZCS on the secondary side, and allows for efficient utilization of the primary side switches with wide changes in load or supply voltage, which is preferred in many applications [2]. The impedance of the series resonant tank is determined by the value of the inductor and capacitor together with the switching frequency. Therefore, it can operate at a higher frequency under a high power level [3] than just dual active bridge topologies without a resonant capacitor, which allows more compact magnetic designs.

The operating modes and controls principles of the DB SRC converters have been extensively analyzed in [5, 6, 7]. Most of the previous works concern the control of the series resonant converter loaded by a diode bridge.

One of the earliest approaches to the control of SRC is based on state-plane trajectory analysis pioneered in the seminal work of R. Oruganti [8]. The key idea is to represent the converter dynamics in the two-dimensional plane of resonant inductor current and resonant capacitor voltage. The analysis is greatly simplified by the fact that with a proper normalization the state plane trajectory of voltage step response of an undamped LC circuit is circular and is centered at the DC solution of the circuit. Thus, accurate and purely geometric analysis can be performed for arbitrary operating point and switching pattern. The downside of this method is that the closed form solutions are becoming very complicated with less trivial PWM patterns. 

The examples of implementation of this strategy [9, 10] for a corresponding class of SRC converters are very simple circuits consist of few analog linear components, logic gates, and flip-flops. Since the control circuit directly drives transistors of the input MOSFET bridge without a PWM modulator these circuits provide excellent transient performance. 
Some recent developments following the geometric state-plane trajectory analysis are presented in [11, 12].
The disadvantage of the approach as mentioned that it is developed only for converters without modulated secondary side, so the voltage gain is always below 1 (i.e. buck mode).

Very closely related to the state-plane trajectory approach are self-sustained oscillating [13, 14] and resonant capacitor charge bang-bang control [15]. Both of these approaches formulate the control problem of converter output voltage using a quantity that directly affects the switching of the input bridge. For the self-sustained oscillating, the voltage controller controls a delay from positive zero crossing of resonant current to the rising edge of input bridge voltage. The capacitor charge bang-bang control uses the resonant capacitor voltage threshold as a control variable and switches the bridge by reaching a threshold of it, defined by the output voltage controller. Both of these techniques suffer from the same drawback as they able to control only the input bridge, and the output secondary side bridge acts as an uncontrollable rectifier.  

The nonlinear control problem of SRC has inevitable nonlinearities and intrinsic constraints. In power supply applications, the controller should be fast enough to reject load transients. These problems are recognized by control systems researchers, and some advanced techniques as passivity-based controllers [16, 17], hybrid-flatness approach [18], and backstepping control approach [19, 20] are presented. All of the works [16-20] consider SRC converters without modulated secondary side, thus the same voltage gain limitation applies.

In recent works, the advantages of the controlled rectifier stage were fully understood and appreciated. The paper [21] introduced and analyzed a new serial resonant converter derived from the classical SRC, with controlled switches at the rectifier stage. The possibility of unlocking voltage gain higher than 1 is emphasized. 

In [22] the bridge shift phase is varied to keep the average power factor at its peak meanwhile the frequency modulation is used to manipulate the power to be transferred. Both bridges are fully driven, no duty cycle control is used, the phase shift between bridges is a function of a measured voltage gain.

A detailed time-domain analysis of ZVS conditions is presented in [23, 24], based on which the closed-form solution is formulated which guarantees sufficient ZVS commutation current at switching instants, accounting for the presence of switches parasitic and/or snubber capacitances, while maintains the RMS tank current at near-to-minimum levels. The phase shift quantities are open-loop based model calculations, without measuring the actual resonant tank current and its position with respect to the bridge switching events.

A similar to [23, 24] control approach is presented in [25], which is based on a detailed model with the multi-harmonics approximation of switching voltage waveforms by Fourier series. The phase shift between bridges is also an open-loop quantity based on the model calculations. Both bridges are fully driven without duty cycle control.

In the paper [26] the resonant tank current is being measured, but a very simple analog controller is proposed for the phase shift between bridges. Also, both bridges are fully driven and the duty cycle control is not explored.

Based on previous publications, the following conclusions could be drawn:

- An attention to control problem of DB SRC is only recently given following appreciation the advantages of this topologies. Most of the published works are remain for SRC converter without a controlled secondary side.

- The duty cycle control in conjunction with phase shift control between bridges is not sufficiently reflected in the published works.

- A closed-loop control of the duty cycles and phase shift between bridges for maintaining ZVS and optimizing efficiency is never considered so far.

In this work, we will systematically analyze the control problem of DB SRC in quasi-steady-state mode for efficiency optimization and output current regulation. We are going to show that by using a simple resonant current zero-crossing detector and measuring timing between resonant tank current and bridge switching events, it is possible to actively control duty cycles of the bridges and the phase shift between primary and secondary sides. This ensures ZVS and maximum efficiency in all operating conditions subject to the model uncertainties.

Our approach of directly inverting the static nonlinearity is technically inspired by [16, 17]. A usage of model-based feedforward maps is based on previous works [23-26], which are augmented by closed loop controllers in the paper. While an idea using the resonant tank zero crossing events is based on [26] and [28].

\section{Background}

This section is a recall of previously published results, and the model [29] in particularly.

\subsection{The converter topology}

\begin{figure}
\centering
\includegraphics[width=1.0\textwidth]{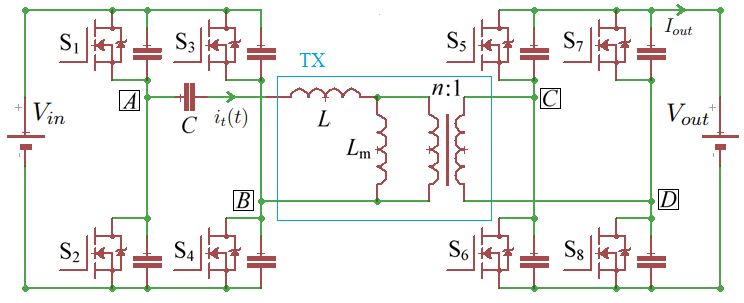}
\caption{DC/DC circuit.}
\label{fig:InitialCircuit}
\end{figure}

The basic electrical circuit of the series LC resonant DC/DC converter is shown in Figure \ref{fig:InitialCircuit}.

The circuit consists of two full (H) transistor bridges: input bridge with switches $S_1$--$S_4$ and output bridge with switches $S_5$--$S_8$. The output bridge is directly connected to the secondary side of transformer TX. The input bridge is connected to the transformer TX through capacitor C. Simple representation of transformer is used with magnetizing inductance $L_m$, leakage inductance $L$ and an ideal transformer with turns ratio $n$. 

If the magnetizing inductance $L_m$ is much larger than leakage inductance $L$: $L_m \gg L$, then there is almost no circulating magnetizing current in the circuit. This has an advantage because all the current from primary side is flowing to the secondary side which increases the efficiency. Also the current in secondary side of transformer is in phase with primary side, which allows easy synchronous rectification by measuring current only in primary side.

By eliminating $L_m$ from the circuit, we can spot that the stray inductance $L$ together with capacitor $C$ forms LC resonant tank. Also ideal transformer can be embedded into secondary bridge for simplification of analysis. Thus we can obtain a black-box circuit presented in Figure \ref{fig:InitialCircuit}. 

The alternating current which flows through the leakage inductance and resonant capacitor is called tank current $i_t(t)$.
This current is being rectified by secondary transistor bridge and assuming that all the ripple components are blocked by output filter capacitors (not shown in Figure \ref{fig:InitialCircuit}), a direct current $I_{out}$ is induced through a load.

\subsection{The voltage waveforms}

Let's consider voltages $v_{in}(t)$ and $v_{out}(t)$ at output terminals of bridges in the circuit \ref{fig:InitialCircuit}. 

\begin{itemize}

\item The voltage $v_{in}(t)$ corresponds to a voltage measured between points A--B of input bridge. 

\item The voltage $v_{out}(t)$ corresponds to scaled by $n$ voltage measured between points C--D of output bridge. 

\end{itemize}

\begin{figure}
\centering
\includegraphics[width=0.8\textwidth]{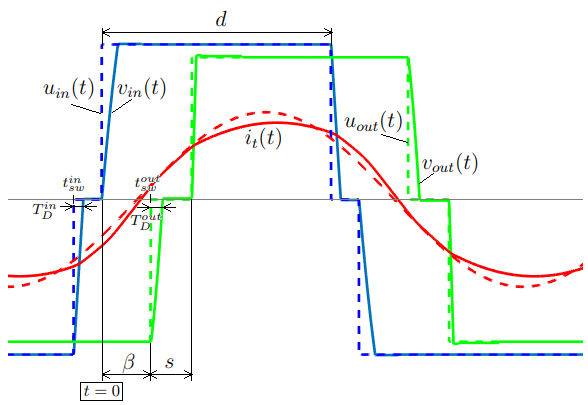}

\caption{Real and approximated voltages applied to resonant LC circuit and PWM parameters.}
\label{fig:Waveforms}
\end{figure}

Effectively, these voltage $v_{in} - v_{out}$ is applied to LC-circuit. 
Figure \ref{fig:Waveforms} shows the real and rectangular approximated voltages.

Let’s consider rectangular approximated voltages $u_{in}(t)$ and $u_{out}(t)$ for easier explanation of controllable switching parameters:

\begin{itemize}

\item The input voltage source is producing rectangular pulses with controlled duty cycle and frequency. The amplitude of pulse is $V_{in}$ and the on-time is $d$ (in radians, i.e. $d = \pi$ is full square wave). The angular frequency is $\omega$, which is equivalent to frequency $F = 1/T$ in Hz.

\item The output voltage source is producing rectangular pulses with amplitude $n V_{out}$ with controlled duty cycle, frequency, and phase shift respect to input bridge. The reciprocal of duty cycle, the off-time or short time is $s$ (defined in radians, $s = \pi$ means that secondary side of transformer is fully shorted). The off-time is always located at the beginning of switching cycle. The phase shift between the output bridge switching cycle and input bridge switching cycle is $\beta$.

\end{itemize}

The output to input voltage ratio is defined as converter voltage gain:

\begin{equation}
G = \frac{n V_{out}}{V_{in}}
\end{equation}

The real waveforms are looking differently than approximated ones because of dead times, which are smoothing rising and falling edges of the switching pulses. Since this DC/DC belongs to the class of resonant converters, energy is also transferring during the dead time, and the dead time is an essential phase of converter operation.

The duration of dead-time for input bridge is $T_D^{in}$, and the duration of dead-time for output bridge is $T_D^{out}$.

\subsection{The model}

The converter model obtained in [29] by first harmonic approximation of tank current can be summarized as:

\begin{equation}\label{eq:affine_model}
W = \frac{I_{out}}{V_{in}} = \frac{n}{2 \pi^2} \frac{\sqrt{A^2 + B^2}}{Z(\omega)} (\cos(s + \delta) + \cos \delta) \\
\end{equation}

where coefficients $A$ and $B$ along with timing quantities $\sigma$ and $\delta$ are given as

\begin{equation}\label{eq:FHA_coeffs}
\begin{gathered}
A = 4 \sin d + 4 G \sin(\beta + s) + 4 G \sin \beta \\
B = 4 - 4 G \cos(\beta + s) - 4 G \cos \beta - 4 \cos d \\
\sigma = \atantwo(B,A), \; \; \delta = \beta - \sigma
\end{gathered}
\end{equation}

and $Z(\omega)$ is resonant LC tank impedance:

\begin{equation}
Z(\omega) = X_L - X_C = \omega L - \frac{1}{\omega C}
\end{equation}

\begin{figure}
\centering
\includegraphics[width=0.8\textwidth]{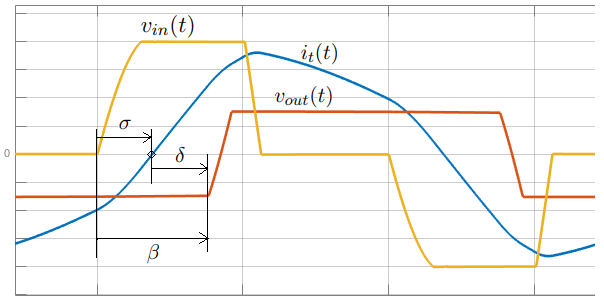}

\caption{Timing of input and output voltages respect to tank current zero level crossing.}
\label{fig:WaveformSync}
\end{figure}

The timing quantities $\sigma$, $\delta$ and $\beta$ are shown in Figure \ref{fig:WaveformSync}

Namely:

\begin{itemize}

\item The $\sigma$ is an angular time (in radians) from the rising edge of positive input bridge voltage $v_{in}(t)$ to a moment when the tank current crosses 0 level: $i_t = 0$. I.e. formally, $i_t(\sigma/\omega) = 0$.

\item The $\delta$ is an angular time (in radians) from the moment when the tank current crosses 0 level $i_t = 0$ to the rising edge of positive output bridge voltage $v_{out}(t)$.

\end{itemize}

It is pretty obvious from the Figure \ref{fig:WaveformSync} that:

\begin{equation}\label{eq:beta}
\sigma + \delta = \beta
\end{equation}

Worth noting that both $\sigma$ and $\delta$ are signed.

\subsection{Regulation problem and optimality metrics}\label{sec:optimality}

In this particular work the DC/DC converter will be represented as a current source. This particular representation covers EV battery charger applications, which is main application area of such topologies.

Thus, an output DC current problem regulation can be formulated as follows:

\textit{Find PWM switching parameters: $d$, $s$, $\beta$, $\omega$ in order to achieve desired steady-state output current:}

\begin{equation}
I_{out}(d,s,\beta,\omega) = I_{out}^*
\end{equation}

where $I_{out}^*$ is reference desired value of output current.

Sine the converter topology considered here is lossless, some additional facts need to be considered for derivation of optimality conditions.
The following requirements can be fulfilled in order to operate the converter efficiently:

\begin{itemize}

\item Conduction loss minimization. 

In order to minimize conduction loss, the amplitude (or RMS) value $I_t$ of tank current $i_t(t)$ needs to be minimized. Thus every set of switching parameters $(d,s,\beta,\omega)$ can be evaluated for conduction loss optimality:

\begin{equation}
\frac{I_t}{I_{out}} \to \min
\end{equation}

\item Turn-on switching loss minimization. 
 
The topology under consideration can be operating in soft-switching mode, which minimizes turn-on losses down to zero. The conditions for soft-switching discussed in many studies. Necessary conditions can be summarized in two statements:

1. A rising edge of positive $v_{in}(t)$ should be at negative tank current $i_t(t)$.

2. A rising edge of positive $v_{out}(t)$ should be at positive tank current $i_t(t)$.

These two statements can be summarized with respect to the timing of switching events. 
Let’s assume that $t = 0$ during positive rising edge of $v_{in}(t)$. Then the soft switching conditions for turn-off are:

\begin{equation}
i_t(0) \le 0, \; \; i_t(\beta/\omega) \ge 0
\end{equation}

or using angular quantities defined in Figure \ref{fig:WaveformSync}:

\begin{equation}\label{eq:sigma_delta_intro}
\sigma \ge 0, \; \; \delta \ge 0
\end{equation}

\item Turn-off switching loss minimization. 

For the turn-off losses, a true soft-switching can't be achieved in this topology. The turn-off losses can be only mitigated. One practical approach is to use a capacitive snubber across the MOSFETs. Additionally, turning-off close to zero current, i.e. achieving ZCS commutation is usually helps achieve better efficiency. 

Thus, we can introduce the following criteria:

\begin{equation}
|i_t(0)| \to \min, \; \; |i_t(\beta/\omega)| \to \min
\end{equation}

It worth noting, that commutating at zero current, i.e. achieving an obvious goal $i_t(0) = i_t(\beta/\omega) = 0$ is not practical since ZVS can be achieved only at specific current levels which allow discharge of internal Coss and external snubber capacitors (if any) during a dead-time interval. 

So generally speaking, the following constraints need to be satisfied:

\begin{equation}
i_t(0) = -I_{th}^{in}, \; \; i_t(\beta/\omega) = I_{th}^{out}
\end{equation}

where $I_{th}^{in}, I_{th}^{out} \ge 0$ are threshold currents for commutation of input and output bridges correspondingly.  

\end{itemize}

\subsection{The output control problem reformulation}

Let's follow the discussion from section \ref{sec:optimality} by taking into account model equations \eqref{eq:affine_model}.

First, achieving the desired output current $I_{out}^*$ is the same as regulating to a prescribed transconductance $W^* = I_{out}^* / V_{in}$ for a given input voltage $V_{in}$. So the reformulated output control problem is to find switching parameters $d$, $s$, $\beta$, $\omega$ so that:

\begin{equation}
W(d,s,\beta,\omega) = W^*
\end{equation} 

where function $W(d,s,\beta,\omega)$ is given by \eqref{eq:affine_model}.

Additionally, from [29] it is known that the amplitude of tank current is given as:

\begin{equation}\label{eq:I_t}
I_t = \frac{V_{in}}{2 \pi Z} \sqrt{A^2 + B^2}
\end{equation}

Using this, the ratio of $I_t/I_{out}$ can be expressed as:

\begin{equation}
\frac{I_t}{I_{out}} = \frac{\pi}{n} \cdot \frac{1}{\cos(s + \delta) + \cos \delta} 
\end{equation}

The ratio $I_t/I_{out}$ is minimized when denominator term $\cos(s + \delta) + \cos \delta$ is maximized. Its maximal value 2 is achieved when both $\delta = 0$ and $s = 0$. 

So the optimality for tank current amplitude can be formulated as follows:

\begin{equation}
\frac{I_t}{I_{out}} \to \min \; \Longleftrightarrow \; \begin{cases} \delta \to \min \\ s \to \min \end{cases}
\end{equation}

This result can be physically interpreted as minimization of reactive power in the resonant tank. 
Furthermore, the case of $\delta = 0$ corresponds to ideal synchronous rectification by the secondary output bridge, i.e. when voltage waveform is fully aligned with tank current.

This type of control which uses phase shift control of the secondary side PWM pulses to align the commutation of the secondary bridge with the sign of the resonant tank is known as \textit{indirect} or PLL-based synchronous rectification [28]. 
In contrast with conventional \textit{direct} synchronous rectification in which the switching of the secondary side is driven by a sign of the resonant tank current detected by a comparator.

The advantages of indirect synchronous rectification to the direct one are twofold:

1. It is possible to compensate comparator's and gate driver's delay by switching the secondary side before the resonant tank current flips polarity, which leads to efficiency increase by minimization of reactive power

2. Since the output of the current sign comparator is no longer driving the secondary side directly, the system is more prone to current measurement noise and potentially can function even without the current sensor (in open loop alignment control, assuming some margins on inefficiency).

\section{Model inversion for commutation parameters $d,s,\beta$}

\subsection{Nonlinear inversion problem formulation}

In order to regulate waveform quantities $\sigma$ and $\delta$ to desired values that maximize efficiency, a control system needs to be implemented around the plant which takes $d,s,\beta$ and outputs $\sigma,\delta$ as measurable quantities.

A feedforward inversion approach is proposed below to linearize $\sigma$ and $\delta$ as functions of controllables $d,s,\beta$.
Thus, the following nonlinear inversion problem can be formulated in form of a system of nonlinear equations:

\begin{equation}\label{eq:inv_problem}
\begin{gathered}
\sigma(d,s,\beta, G) = \sigma^* \\
\delta(d,s,\beta, G) = \delta^*
\end{gathered}
\end{equation}

with respect to unknown variables $d \in [0,\pi]$, $s \in [0, \pi]$, $\beta \in [-\pi, \pi]$ and for given referenced values $\sigma^* \in [-\pi/2, \pi/2]$, $\delta^* \in [-\pi/2, \pi/2]$ and parameter $G \ge 0$.

\begin{figure}[H]
\input{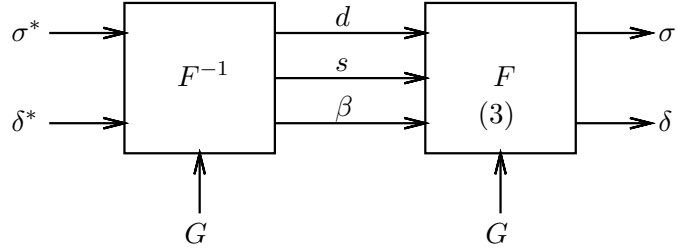}

\caption{Nonlinear inversion with respect to $\sigma$ and $\delta$.}
\label{fig:Inversion}
\end{figure}

The concept of nonlinear inversion can be illustrated by a block diagram in Figure \ref{fig:Inversion}, where nonlinear function $F$ is given by model equations \eqref{eq:FHA_coeffs} and $F^{-1}$ is a solution of \eqref{eq:inv_problem}.

\subsection{Nonlinear inversion condition}\label{sec:inversion_condition}

Let's recall \eqref{eq:FHA_coeffs} as $\sigma^* = \atantwo(B,A)$. In order to have $\sigma^* \in [-\pi/2, \pi/2]$, the argument $A$ should not be negative:

\begin{equation}\label{eq:sigma_domain}
A = \sin d + G \sin(\beta + s) + G \sin \beta \ge 0
\end{equation}

If this holds, then $\atantwo$ can be replaced by $\tan$, and the nonlinear inversion problem can be rewritten as following two equations:

\begin{equation}
\begin{gathered}
\frac{1 - G \cos(\beta + s) - G \cos \beta - \cos d}{\sin d + G \sin(\beta + s) + G \sin \beta} = \tan \sigma^* \\
\delta^* + \sigma^* = \beta 
\end{gathered}
\end{equation}

Note that $\beta = \delta^* + \sigma^*$ became constant. 

This can be transformed as follows:

\begin{equation}
\tan \sigma^* \cdot \left( G \sin(\beta + s) + G \sin \beta + \sin d \right ) + G \cos(\beta + s) + G \cos \beta + \cos d - 1 = 0
\end{equation}

And the equation can be simplified further by using identity of harmonic addition:

\begin{equation}
\tan \sigma^* \sin(x) + \cos(x) = \sqrt{(\tan \sigma^*)^2 + 1} \cdot \cos(x - \atan \tan \sigma^*) = \sqrt{(\tan \sigma^*)^2 + 1} \cdot \cos(x - \sigma^*)
\end{equation}

where $x$ is an arbitrary angular argument.

Furthermore, taking into account the following trigonometric identity:

\begin{equation}
\frac{1}{\sqrt{(\tan \sigma^*)^2 + 1}} = |\cos \sigma^*|
\end{equation}

and using harmonic addition, the nonlinear inversion condition can be rewritten as:

\begin{equation}
G \cos(\beta + s - \sigma^*) + G \cos(\beta - \sigma^*) + \cos(d - \sigma^*) - \cos \sigma^* = 0
\end{equation}

Note that the absolute value of $\cos \sigma^*$ is safely omitted since the range of $\sigma^*$ is $[-\pi/2, \pi/2]$.

Finally, using $\beta = \delta^* + \sigma^*$, it simplifies even further, which we will refer as \textit{nonlinear inversion condition} (for $d$ and $s$):

\begin{equation}\label{eq:inversion_d_s}
G \cos(\delta^* + s) + G \cos \delta^* + \cos(d - \sigma^*) - \cos \sigma^* = 0
\end{equation}

It is important to note that \eqref{eq:inversion_d_s} is underdetermined since it has two unknown variables $d$ and $s$. 
So in order to find a unique solution, the minimization of $s \to \min$ should be additionally imposed.

\subsubsection{Buck mode with $s = 0$}

This is the simplest mode of operation and obviously satisfies minimization condition since the $s = 0$ is minimal value of $s$.
Substituting this to \eqref{eq:inversion_d_s}

\begin{equation}
2 G \cos \delta^* + \cos(d - \sigma^*) - \cos \sigma^* = 0
\end{equation}

thus:

\begin{equation}\label{eq:d_inversion}
d = \acos \left (\cos \sigma^* - 2 G \cos \delta^* \right ) + \sigma^*
\end{equation}

The domain of this buck operating mode is given by domain of $\acos$ function and the range of $d \in [0,\pi]$

\begin{equation}
\begin{gathered}
-1 \le \cos \sigma^* - 2 G \cos \delta^* \le 1 \\
0 \le \acos \left (\cos \sigma^* - 2 G \cos \delta^* \right ) + \sigma^* \le \pi
\end{gathered}
\end{equation}

Practically, since the $\sigma^*$ is additive to result of $\acos$ in \eqref{eq:d_inversion}, it is pretty much possible that the calculation of $d$ could give greater than $\pi$ value. The boundary condition is 

\begin{equation}
\cos \sigma^* - 2 G \cos \delta^* = \cos \left( \pi - \sigma^* \right ) = -\cos \sigma^*
\end{equation}

and the domain for $d \le \pi$ is defined by inequality:

\begin{equation}\label{eq:buck_domain}
\cos \sigma^* \ge G \cos \delta^*
\end{equation}

Another useful boundary case is when the argument of $\acos$ in \eqref{eq:d_inversion} is out of $[-1,1]$ range, particularly for high values of $G$, which corresponds to a boost mode:

\begin{equation}
\cos \sigma^* - 2 G \cos \delta^* = -1
\end{equation}

which translates to the following inequality:

\begin{equation}
\cos \sigma^* \ge 2 G \cos \delta^* - 1
\end{equation}

\subsubsection{Boost mode with $d = \pi$}

Substituting $d = \pi$ to \eqref{eq:inversion_d_s} gives:

\begin{equation}
G \cos(\delta^* + s) + G \cos \delta^* - 2 \cos \sigma^* = 0
\end{equation}

thus for variable $s$ it gives:

\begin{equation}\label{eq:s_inversion}
s = \acos \left( 2 \cos \sigma^* / G - \cos \delta^* \right ) - \delta^*
\end{equation}

The domain of this boost operating mode is given by domain of $\acos$ function and the range of $s \in [0,\pi]$

\begin{equation}
\begin{gathered}
-1 \le 2 \cos \sigma^* / G - \cos \delta^* \le 1 \\
0 \le \acos \left( 2 \cos \sigma^* / G - \cos \delta^* \right ) - \delta^* \le \pi
\end{gathered}
\end{equation}

Practically, since the $\delta^*$ is additive to result of $\acos$ in \eqref{eq:s_inversion}, it is pretty much possible that the calculation of $s$ could give negative value. The boundary condition is 

\begin{equation}
2 \cos \sigma^* / G = 2 \cos \delta^*
\end{equation}

and the domain for $s \ge 0$ is defined by inequality:

\begin{equation}
\cos \sigma^* \le G \cos \delta^*
\end{equation}

Another useful boundary case is when the argument of $\acos$ in \eqref{eq:d_inversion} is out of $[-1,1]$ range, particularly for low values of $G$, which corresponds to a buck mode:

\begin{equation}
2 \cos \sigma^* / G - \cos \delta^* = 1
\end{equation}

which translates to the following inequality:

\begin{equation}
2 \cos \sigma^* / G \le 1 + \cos \delta^*
\end{equation}

\subsubsection{Explicit additive reference of $s$}\label{sec:s_add}

There are use cases when both $d < \pi$ and $s > 0$ are required. In this case, value of $s$ consists of two parts: a value required by \eqref{eq:s_inversion} which will be denoted as $s_{min}$ and an additive component $s_{add} \ge 0$ which is specified externally:

\begin{equation}
s = s_{min} + s_{add}
\end{equation}

It worth noting that just increasing $s$ by addition of $s_{min}$ without corresponding adjustment of $d$ and $\beta$ will change resulting $\sigma$ and $\delta$. So in order to maintain prescribed $\sigma^*$ and $\delta^*$, the $d$ and $\beta$ need to be adjusted.

Thus, the nonlinear inversion equation \eqref{eq:inversion_d_s} needs to be modified as follows:

\begin{equation}\label{eq:inversion_d_s__s_add_}
G \cos(\delta^* + s_{min} + s_{add}) + G \cos \delta^* + \cos(d - \sigma^*) - \cos \sigma^* = 0
\end{equation}

where $s_{add}$ is a constant.

First consider the case of $s_{min} = 0$, which corresponds to the buck mode.

\begin{equation}
d = \acos \left ( \cos \sigma^* - G \cos(\delta^* + s_{add}) - G \cos \delta^* \right ) + \sigma^*
\end{equation}


The buck mode is defined by a proper range of $d \le \pi$ when no $s$ is needed except the $s_{add}$ commanded externally, so that $s_{min} = 0$, $s = s_{add}$. Thus it is possible to generalize \eqref{eq:buck_domain} as follows:

\begin{equation}
2 \cos \sigma \ge G \cos(\delta + s_{add}) + G \cos \delta
\end{equation}

If the buck mode with $d \le \pi$ and $s = s_{add}$ is not feasible, then the boost mode is used.
A minimal value of $s$ without $s_{add}$ is given by \eqref{eq:s_inversion} which is

\begin{equation}\label{eq:s_min}
s_{min} = \acos \left( 2 \cos \sigma^* / G - \cos \delta^* \right ) - \delta^*
\end{equation}

such that $s$ is sum of $s_{min}$ and $s_{add}$:

\begin{equation}\label{eq:s_inversion__s_add}
s = \acos \left( 2 \cos \sigma^* / G - \cos \delta^* \right ) - \delta^* + s_{add}
\end{equation}

Then in presence of $s_{add} > 0$, the value of $d$ which satisfies inversion equation \eqref{eq:inversion_d_s__s_add_} needs to be calculated as follows:

\begin{equation}\label{eq:d_inversion__s_add}
d = \acos \left ( \cos \sigma^* - G \cos(\delta^* + s_{min} + s_{add}) - G \cos \delta^* \right ) + \sigma^*
\end{equation}

The steps for obtaining the inversion map $F^{-1}$ are: 

- Check condition \eqref{eq:buck_domain}, if it holds, then this is a buck mode, use $s_{min} = 0$ and $s = s_{add}$.

- If \eqref{eq:buck_domain} doesn't hold, then this is a boost mode and $s$ is calculated by \eqref{eq:s_inversion__s_add}.

- Calculate $d$ using \eqref{eq:d_inversion__s_add}.

- Check inequality \eqref{eq:sigma_domain}. If it is not satisfied, then the provided $\delta^*$ and $\sigma^*$ are not feasible.

\subsection{Examples of feed-forward characteristics}

The solution \eqref{eq:d_inversion__s_add},\eqref{eq:s_inversion__s_add} of the inversion problem \eqref{eq:inv_problem} is evaluated for different $\sigma, \delta$ and $G$. The results are presented in this section in form of contour plots.

\subsubsection{Buck mode $G = 0.5$}

\begin{figure}[H]
\centering
\includegraphics[width=1.1\textwidth]{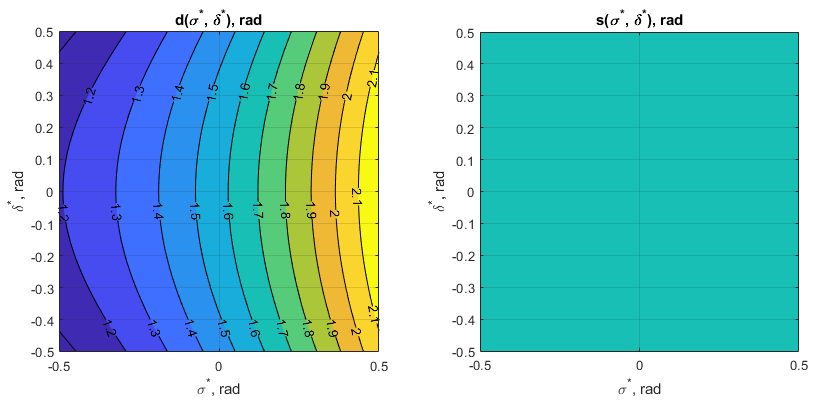}

\caption{The switching waveform parameters $d$ and $s$ as functions of required $\sigma$ and $\delta$ for buck mode $G = 0.5$ and $s_{add} = 0$ (resulting in $s=0$)}
\end{figure}

\begin{figure}[H]
\centering
\includegraphics[width=1.1\textwidth]{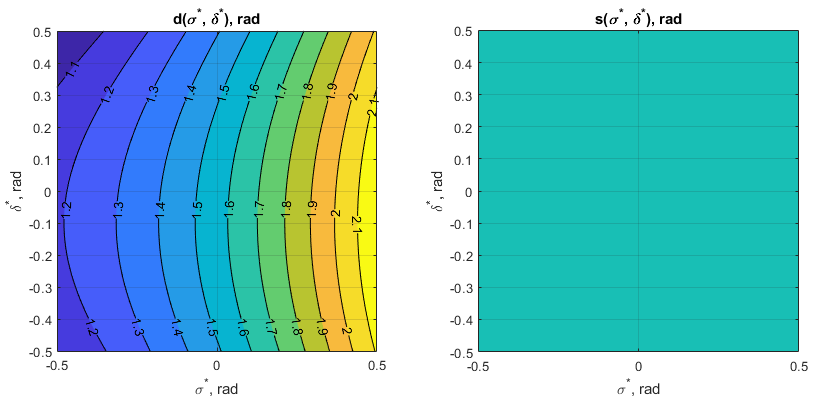}

\caption{The switching waveform parameters $d$ and $s$ as functions of required $\sigma$ and $\delta$ for buck mode $G = 0.5$ and $s_{add} = 0.2$ (resulting in $s=0$)}
\end{figure}

\begin{figure}[H]
\centering
\includegraphics[width=1.1\textwidth]{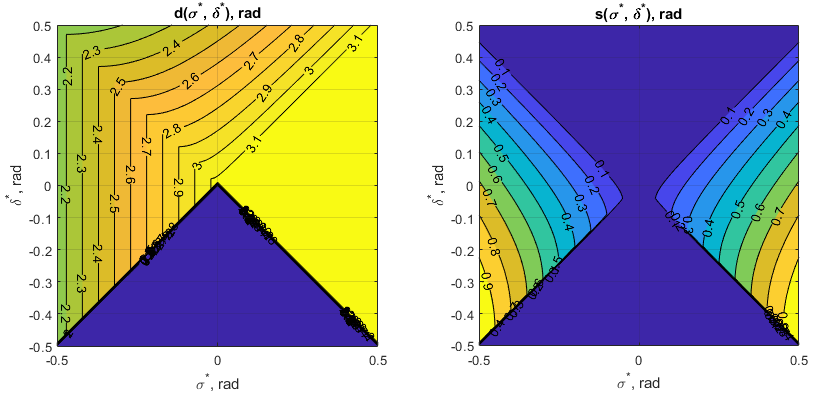}

\caption{The switching waveform parameters $d$ and $s$ as functions of required $\sigma$ and $\delta$ for $G = 1$ and $s_{add} = 0$}
\end{figure}

\begin{figure}[H]
\centering
\includegraphics[width=1.1\textwidth]{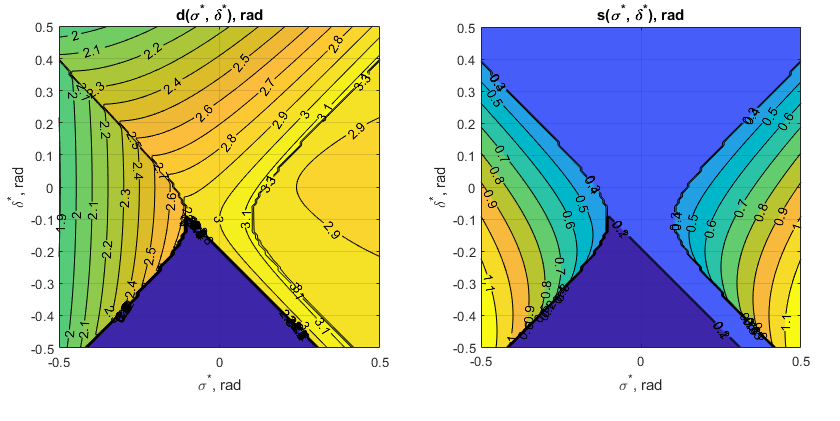}

\caption{The switching waveform parameters $d$ and $s$ as functions of required $\sigma$ and $\delta$ for $G = 1$ and $s_{add} = 0.2$}
\end{figure}

\begin{figure}[H]
\centering
\includegraphics[width=1.1\textwidth]{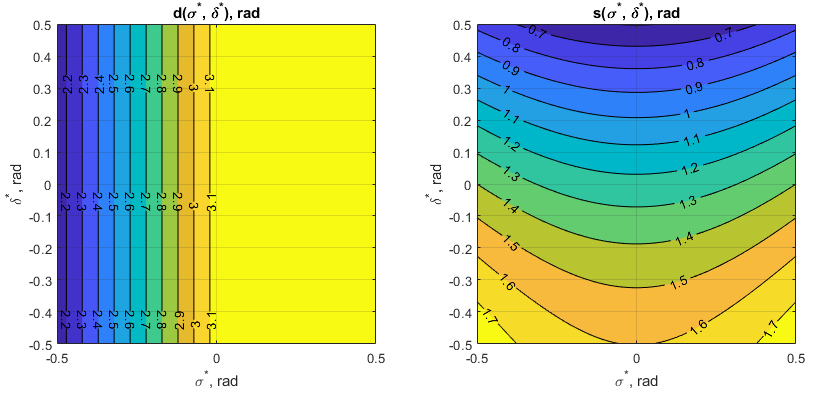}

\caption{The switching waveform parameters $d$ and $s$ as functions of required $\sigma$ and $\delta$ for boost mode $G = 1.5$ and $s_{add} = 0$}
\end{figure}

\begin{figure}[H]
\centering
\includegraphics[width=1.1\textwidth]{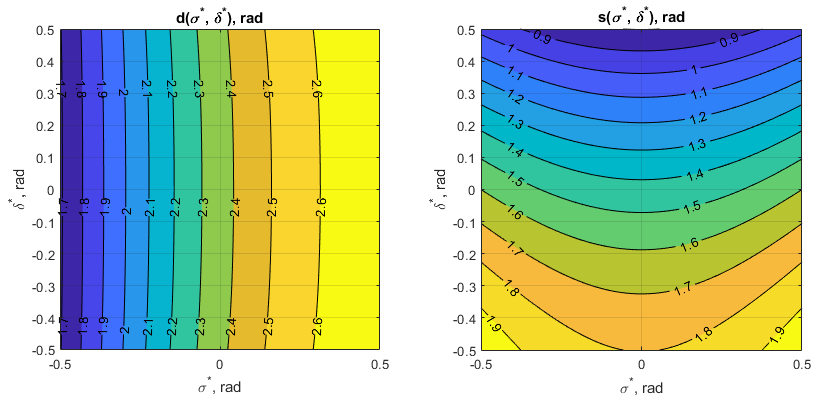}

\caption{The switching waveform parameters $d$ and $s$ as functions of required $\sigma$ and $\delta$ for boost mode $G = 1.5$ and $s_{add} = 0.2$}
\end{figure}

\subsection{Simulation of the feed-forward control}\label{sec:ff_sim}

The proposed feed-forward control \eqref{eq:s_inversion__s_add}, \eqref{eq:d_inversion__s_add} is simulating along with the plant steady-state model \eqref{eq:FHA_coeffs} in order to illustrate the control trajectories of PWM quantities $d,s,\beta$.
In this simulation a sinisouidal reference trajectories are generated for $\sigma^*, \delta^*, s_{add}$. The $G$ swept from 0 to 2 in time so that $G = t$.

The results are presented in Figures below

\begin{figure}[H]
\centering
\includegraphics[width=0.7\textwidth]{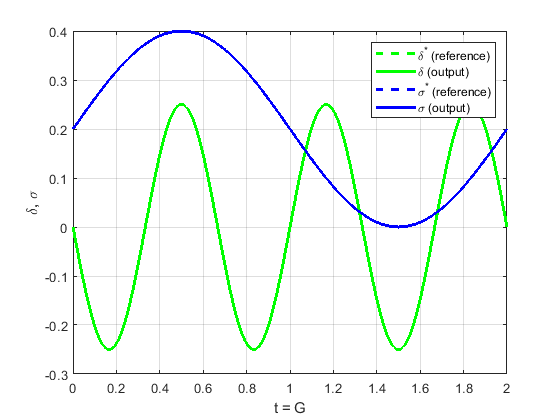}

\caption{Simulated trajectories of $\sigma$ and $\delta$: corresponding reference and output curves are overlapping}
\end{figure}

\begin{figure}[H]
\centering
\includegraphics[width=0.7\textwidth]{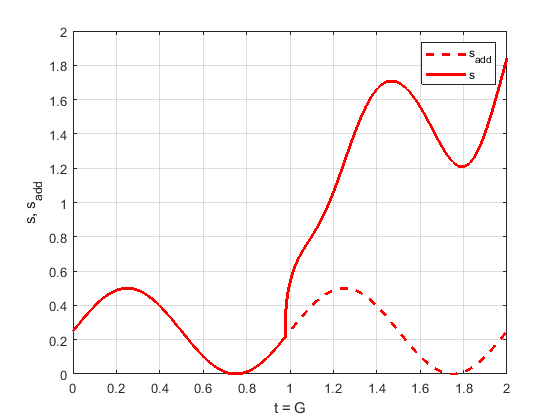}

\caption{Simulated trajectories of $s$ and $s_{add}$: $s_{add} = s$ for buck mode $G < 1$}
\end{figure}

\begin{figure}[H]
\centering
\includegraphics[width=0.7\textwidth]{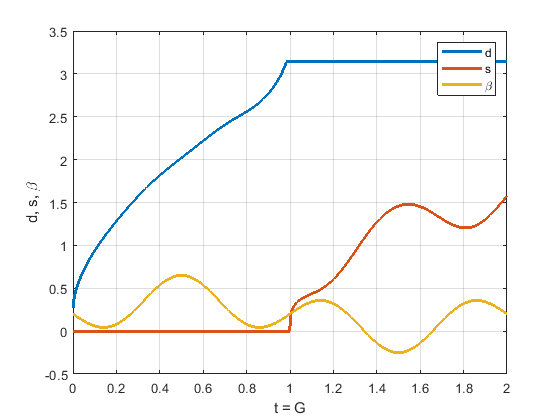}

\caption{Simulated trajectories of control inputs $s,d,\beta$}
\end{figure}

\subsection{Special cases}

\subsubsection{Special case of $G = 1$}\label{sec:G_1}

It is important to discuss case of $G = 1$, because this operating point has unique properties. Furthermore, it is possible to show that $G = 1$ is most efficient operating point with respect to the active losses in the circuit.

Substituing $G = 1$, $s_{add} = 0$ and $\delta^* = 0$ to \eqref{eq:d_inversion__s_add} and \eqref{eq:s_inversion__s_add} gives:

\begin{equation}
\begin{gathered}
s = \acos \left( 2 \cos \sigma^* - 1 \right ) \\
d = \acos \left ( \cos \sigma^* - 2 \right ) + \sigma^*
\end{gathered}
\end{equation}

It is pretty obvious that $s = 0$ and $d = \pi$ for $\sigma^* = 0$, along with $\beta = 0$.  
Then calculating coefficients $A$ and $B$ by \eqref{eq:FHA_coeffs} gives: $A = 0$ and $B = 0$. Thus, the amplitude of tank current is $I_t = 0$ according to \eqref{eq:I_t}. This leads to an important conclusion: in order to prevent collapse of tank current amplitude, the refernce value of $\sigma^*$ should be non-zero: $\sigma^* > 0$.

\subsubsection{Ideal synchronous rectification with $\delta = 0$}

Let's consider second equation in \eqref{eq:affine_model}. In order to satisfy the control goal $\delta = 0$ for synchronous rectification, the following should be true:

\begin{equation}
\delta = \beta - \sigma = 0
\end{equation}

which gives trivially

\begin{equation}
\beta = \sigma
\end{equation}

or

\begin{equation}
\tan \beta = \tan \sigma
\end{equation}

or expanding by using \eqref{eq:FHA_coeffs} and considering only $\tan \beta$ in the left hand side:

\begin{equation}
\frac{\sin \beta}{\cos \beta} = \frac{B}{A} = \frac{1 - G \cos(\beta + s) - G \cos \beta - \cos d}{\sin d + G \sin(\beta + s) + G \sin \beta}
\end{equation}

After some trivial simplification, the last equation can be simplified down to the following, which we will call \textit{synchronous rectification condition}:

\begin{equation}\label{eq:sync_rect}
\cos \beta - G - \cos (\beta - d) - G \cos s = 0
\end{equation}

\subsubsection{Edge case of $\beta = 0$}

As discussed earlier, it is beneficial to bring primary and secondary voltage waveforms close to each other in order to minimize commutation current. Such approach is not always practical since the current might be not enough to ensure ZVS commutation. However, this case is interesting in terms of theoretical analysis of converter characteristics.

Substituting $\beta = 0$ into \eqref{eq:sync_rect} immediately gives:

\begin{equation}\label{eq:no_beta}
1 - G - \cos d - G \cos s = 0
\end{equation}
 
Since from the previous section, $\sigma = \beta$ then the equation \eqref{eq:no_beta} describes a case when both input and output waveforms are brought together.

\begin{figure}
\centering
\includegraphics[width=0.9\textwidth]{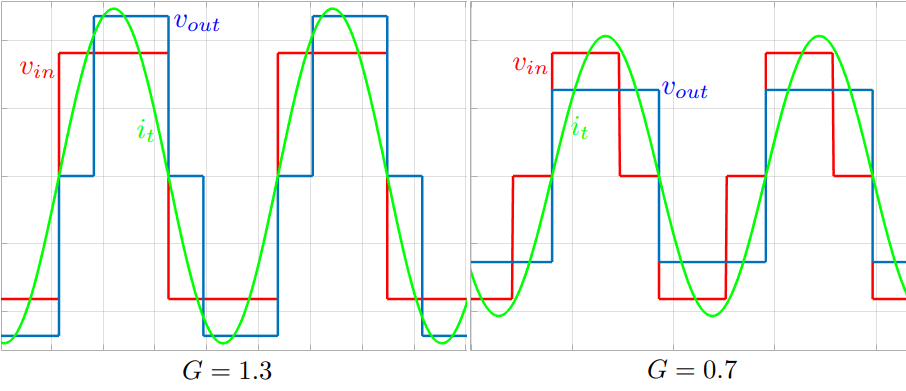}

\caption{Voltage and current waveforms aligned with $\beta = 0$ for buck ($G = 0.7$) and boost ($G = 1.3$) voltage ratios (under the same input voltage and frequency)}
\label{fig:WaveformsBeta0BuckBoost}
\end{figure}

It is interesting to study domain of arguments for \eqref{eq:no_beta}. The equation can be resolved with respect to $d$:

\begin{equation}\label{eq:d__Gs}
d = \acos(1 - G - G \cos s)
\end{equation}

and in the same time can be resolved with resprct to $s$:

\begin{equation}\label{eq:s__Gd}
s = \acos \left ( \frac{1  - \cos d}{G} - 1 \right )
\end{equation}

Lets consider two cases:

\begin{itemize}

\item $s = 0$

From \eqref{eq:d__Gs}, the following is evident:

\begin{equation}\label{eq:d__ctrl}
d = \acos(1 - 2 G)
\end{equation}

which is defined on interval $0 \ge G \ge 1$. 

If $G = 1$, then $d = \pi$. 

It is also easy to see that when $d = \pi$ or $1 - G - G \cos s = -1$, then $\cos s = 2/G - 1$ and $s$ increases with the increase of $G \ge 1$.

\item $d = \pi$

From \eqref{eq:s__Gd}, the following is evident:

\begin{equation}\label{eq:s__ctrl}
s = \acos \left ( \frac{2}{G} - 1 \right )
\end{equation}

which is defined on interval $G \ge 1$. 

\end{itemize}

\begin{figure}
\centering
\begin{tikzpicture}
  \begin{axis}[ 
    width=0.8*\textwidth,
    height=\axisdefaultheight,
    xlabel=$G$,
    ylabel={$d, \; s$, rad},
    samples=100,
    domain=0:2,
    grid=both
  ] 
    \addplot[blue,no marks,very thick] { pi*acos(1 - 2*min(x, 1))/180 }; 
    \addlegendentry{$d(G)$} 
    \addplot[red,no marks,very thick] { pi*acos( 2 / max(x, 1) - 1 )/180 }; 
    \addlegendentry{$s(G)$}
  \end{axis}
\end{tikzpicture}

\caption{Maps of $d(G)$ and $s(G)$}
\label{fig:d_s__G}
\end{figure}
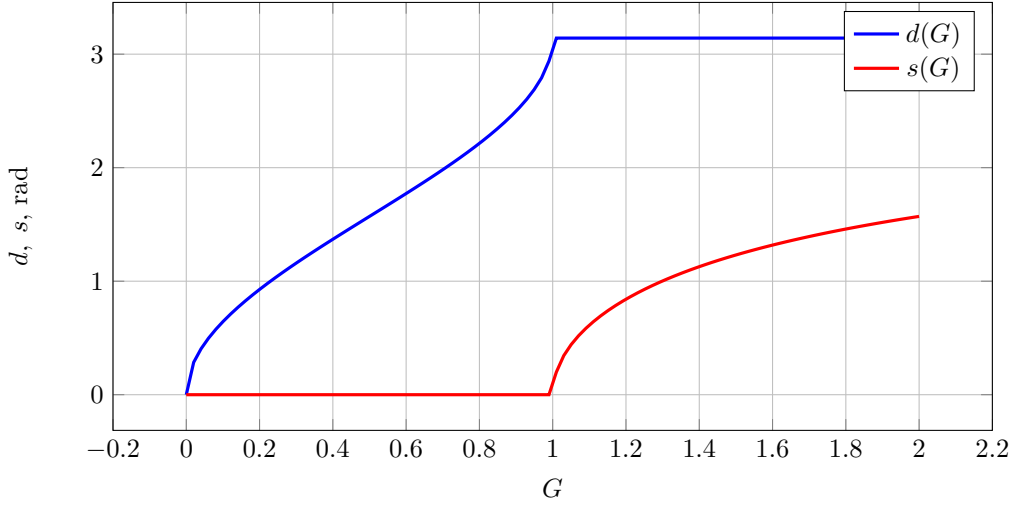

The combined characteristics of $d(G)$ and $s(G)$ are shown in Figure \ref{fig:d_s__G}

It is possible to combine both feedforward control laws \eqref{eq:d__ctrl} and \eqref{eq:s__ctrl} into the single one by introducing a new artificial variable $q$ piecewise parametrized by $G$ and defined as follows:

\begin{equation}\label{eq:q}
q(d,s) = \begin{cases} d, \; d < \pi \\ s + \pi, \; d = \pi \end{cases} = \begin{cases} d, \; G \le 1 \\ s + \pi, \; G > 1 \end{cases}
\end{equation}

with the following inverse mapping back to $d$ and $s$:

\begin{equation}
d(q) = \begin{cases} q, \; G \le 1 \\ \pi, \; G > 1 \end{cases}, \; s(q) = \begin{cases} 0, \; G \le 1 \\ q - \pi, \; G > 1 \end{cases}
\end{equation}

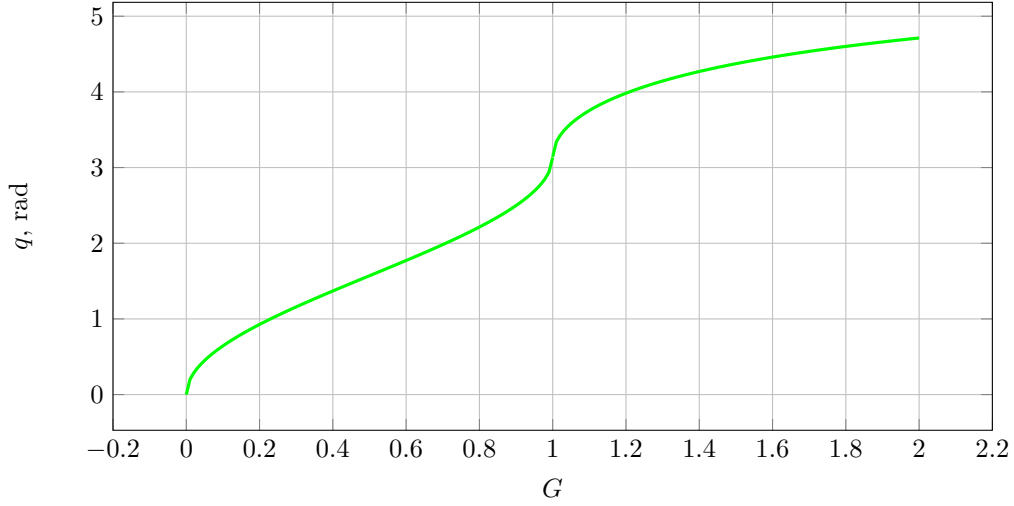
\begin{figure}
\centering
\begin{tikzpicture}
  \begin{axis}[ 
    width=0.8*\textwidth,
    height=\axisdefaultheight,
    xlabel=$G$,
    ylabel={$q$, rad},
    samples=100,
    domain=0:2,
    grid=both
  ] 
    \addplot[green,no marks,very thick][domain=0:1] { pi*acos(1 - 2*x)/180 }; 
    \addplot[green,no marks,very thick][domain=1:2] { pi + pi*acos( 2 / x - 1 )/180 }; 
  \end{axis}
\end{tikzpicture}

\caption{Map of $q(G)$}

\end{figure}

\subsubsection{Primary side fully driven converter}

In two subsections below a special case of the PWM modulation with fully driven primary side will be discussed. The operating mode of primary input bridge is the same as for the conventional LLC converter. However, with a control of secondary side duty cycle $s$ it is possible to obtain a boost operating mode and significantly increase voltage range of the converter.

Technically, driving the converter in this PWM type of modulation will give a sligtly worse efficiency (because the primary bridge is being switched at higher instantenous current), but this type if modulation might be desirable for EMI purposes, because the switching voltage has fewer harmonics.

Everywhere in this section we will assume $d = \pi$.

\paragraph{Buck mode with $d = \pi$ and $s = 0$.}

Lets consider first a buck mode with $s = 0$, which corresponds to a fully driven converter on primary side and a synchronous rectification on the secondary side. Obviously, buck mode is optimal in terms of $s \to \min$, since 0 is minimal possible value of $s$. 

From \eqref{eq:sync_rect} follows that since $\cos s = 1$, then

\begin{equation}\label{eq:buck_beta}
\beta = \sigma = \acos G
\end{equation}

It immediately follows that in order to have a real solution for $\beta$, the voltage ratio should be $G \le 1$, which coincides with the buck mode of converter operation. 

To satisfy control goal constraint $\sigma \ge \sigma_{min}$, additional consideration should be imposed for voltage ratio $G$:

\begin{equation}
\begin{gathered}
\sigma \ge \sigma_{min} \\
\acos G \ge \sigma_{min} \\
G \le \cos ( \sigma_{min} ) 
\end{gathered}
\end{equation}

Thus, as long as $G \le \cos (\sigma_{min})$, the buck mode control with $s = 0$ can satisfy goal constraint $\sigma \ge \sigma_{min}$, and voltage ratio $G = \cos ( \sigma_{min} )$ is a boundary between buck and boost modes of converter operation.

\paragraph{Boost mode with $d = \pi$ and $s > 0$, buck-boost transition.}

The boost mode is activated when the buck mode with $s = 0$ is no longer capable to satisfy the control goal for $\sigma$, and as it was established in previous section, for $G \ge \cos (\sigma_{min})$ a boost mode should be used. In order to satisfy optimality condition $s \to \min$, the $\sigma$ should be fixed in boost mode, and we have following control problem for angles:

\begin{equation}
\begin{gathered}
\delta = 0 \\
\sigma = \sigma_{min}
\end{gathered}
\end{equation}

or 

\begin{equation}\label{eq:boost_beta}
\sigma = \beta = \sigma_{min}
\end{equation}

From \eqref{eq:sync_rect} the shorting time $s$ is determined by:

\begin{equation}\label{eq:boost_s}
s = \acos \left ( 2 \cos( \sigma_{min}) / G - 1 \right )
\end{equation}

Note for the particular voltage ratio $G = \cos (\sigma_{min})$, i.e. exactly at the buck to boost transition border, the $s = 0$ according to \eqref{eq:boost_s}, which proofs continuity of $s(G)$ function. The same is for $\beta$, since $\beta = \acos G$ in buck mode, then at critical voltage ratio $G = \cos (\sigma_{min})$ we are obtaining $\beta = \acos \cos (\sigma_{min}) = \sigma_{min}$, which coincides with continuity of $\beta(G)$ function.

\begin{figure}
\centering
\begin{tikzpicture}
  \begin{axis}[ 
    width=0.8*\textwidth,
    height=\axisdefaultheight,
    xlabel=$G$,
    ylabel={$\beta, \; s$},
    samples=100,
    domain=0:2,
    grid=both
  ] 
    \addplot[blue,no marks,very thick] { pi*acos(min(x, 0.95))/180 };
    \addlegendentry{$\beta(G)$} 
    \addplot[red,no marks,very thick] { pi*acos( 2*0.95 / max(x, 0.95) - 1 )/180 };
    \addlegendentry{$s(G)$}
  \end{axis}
\end{tikzpicture}

\caption{Maps of $\beta(G)$ and $s(G)$ for particular $G^* = 0.95$}
\label{fig:beta_s__G}
\end{figure}
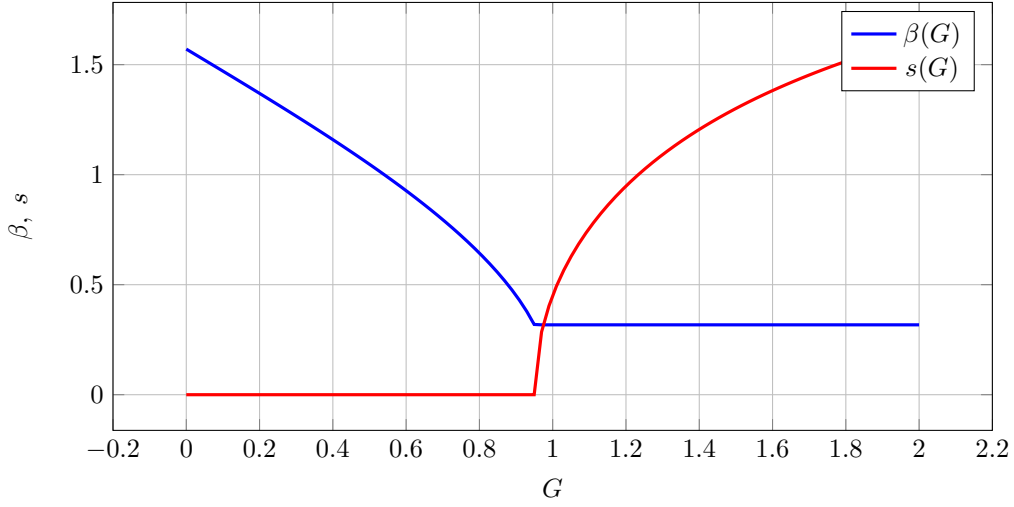

We can combine $\beta$ and $s$ maps for both buck and boost modes by stitching \eqref{eq:buck_beta}, \eqref{eq:boost_beta} and \eqref{eq:boost_s} using minimum and maximum functions (result is pictured in Figure \ref{fig:beta_s__G}):

\begin{equation}
\begin{gathered}
\beta = \acos( \min \{ G, G^* \} ) \\
s = \acos \left ( \frac{2 G^*}{\max \{ G, G^* \}} - 1 \right )
\end{gathered}
\end{equation}

where 

\begin{equation}
G^* = \cos (\sigma_{min})
\end{equation}

\subsubsection{Linearization around $\delta^* = 0$ and $\sigma^* = 0$}

Even though the approach presented in this paper doesnt rely on any kind of linearization (except harmonic truncation for the first harmonic linearization), it is interesting to examine behaviour of the linearization maps around steady-state point of $\delta^* = 0$ and $\sigma^* = 0$.

Lets assume $s_{add} = 0$ for simplicity and lets linearize \eqref{eq:d_inversion} and \eqref{eq:s_inversion} around $\delta^* = 0$ and $\sigma^* = 0$:

\begin{equation}\label{eq:d_linear}
\begin{gathered}
d = d(\delta^*,\sigma^*) \approx d(0,0) +  \frac{\partial d}{\partial \delta} \biggr\rvert_{\delta = 0} \cdot \delta^* + \frac{\partial d}{\partial \sigma} \biggr\rvert_{\sigma = 0} \cdot \sigma^* = \\
= \acos(1 - 2 G) - \frac{G \sin \delta}{\sqrt{G (1-G)}}  \biggr\rvert_{\delta = 0} \cdot \delta^* + \frac{\sin \sigma}{2 \sqrt{G (1-G)}} \biggr\rvert_{\sigma = 0} \cdot \sigma^* + \sigma^*
\end{gathered}
\end{equation}

\begin{equation}\label{eq:s_linear}
\begin{gathered}
s = s(\delta^*,\sigma^*) \approx s(0,0) + \delta^* \frac{\partial s}{\partial \delta} \biggr\rvert_{\delta = 0} \cdot \delta^* + \sigma^* \frac{\partial s}{\partial \sigma} \biggr\rvert_{\sigma = 0} \cdot \sigma^* = \\
=  \pi - \acos(1 - 2/G) - \frac{G \sin \delta}{2 \sqrt{G - 1}} \biggr\rvert_{\delta = 0} \cdot \delta^* + \frac{\sin \sigma}{\sqrt{G - 1}} \biggr\rvert_{\sigma = 0} \cdot \sigma^* - \delta^*
\end{gathered}
\end{equation}

Thus, in trivial case, the decoupling transformation can be approximated as follows:

\begin{equation}\label{eq:ds_linear}
\begin{gathered}
d = \begin{cases} \acos(1 - 2 G) + \sigma^*, \; G < 1 \\ \pi, \; G > 1 \end{cases} \\
s = \begin{cases} 0, \; G < 1 \\ \pi - \acos(1 - 2/G) - \delta^*, \; G > 1 \end{cases} \\
\beta = \sigma^* + \delta^*
\end{gathered}
\end{equation}

This transformation is not defined at $G = 1$, because according to section \ref{sec:G_1}, in order to prevent the tank current from a collapse to zero the following should be true $\sigma^* > 0$. Also according to \eqref{eq:d_linear} and \eqref{eq:s_linear}, partial derivatives $\partial d/\partial \delta$, $\partial d/\partial \sigma$, $\partial s/\partial \delta$, $\partial s/\partial \sigma$ are all singular near $G = 1$.

\begin{figure}[H]
\centering
\includegraphics[width=0.7\textwidth]{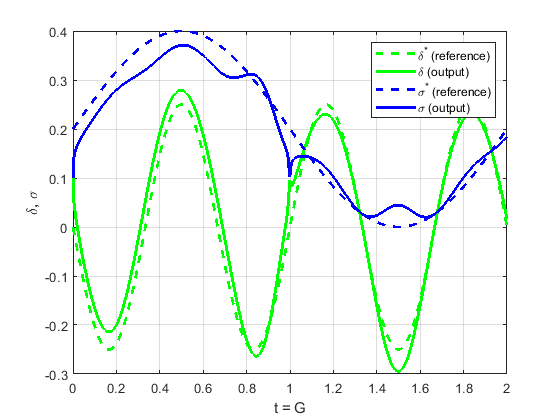}

\caption{Simulated trajectories of $\sigma$ and $\delta$ without external feedback controller, only using linear map \eqref{eq:ds_linear}}
\label{fig:delta_sigma_ctrl_lin}
\end{figure}

\begin{figure}[H]
\centering
\includegraphics[width=0.7\textwidth]{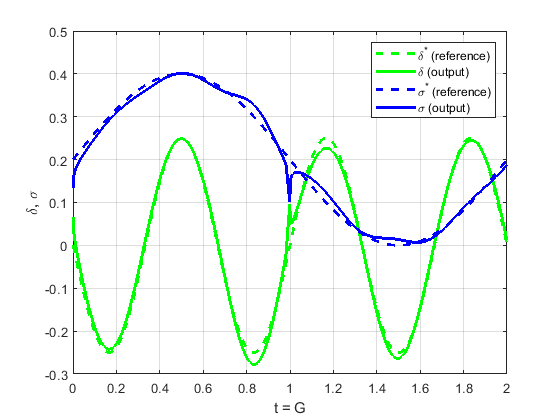}

\caption{Simulated trajectories of $\sigma$ and $\delta$ using individual PI controllers in addition to feed-forward term \eqref{eq:ds_linear}}
\label{fig:delta_sigma_ctrl_PID}
\end{figure}

It worth noting that this \eqref{eq:ds_linear} is only an approximation of exact inversion \eqref{eq:d_inversion} and \eqref{eq:s_inversion}. An external feedback controller has to be used in order to achieve satisfactory results. 
This can be illustrated by simulation results presented in Figures \ref{fig:delta_sigma_ctrl_lin} and \ref{fig:delta_sigma_ctrl_PID}.

The conclusions from the simulation results are: introduction of PI controllers are enhancing regulation accuracy in comparision to purely feed-forward control \eqref{eq:ds_linear}, also the control trajectories have a cusp in $G = 1$, which is expected since the control approximation is discontinuous there.

\section{Output power control}

\subsection{Variable frequency control}

According to \eqref{eq:affine_model} the output current control can be formulated in terms of transconductance $W$ by simply multiplying it by input voltage $V_{in}$: $I_{out} = W V_{in}$. Thus, the output current can be regulated by means of transconductance regulation by satisfying the following control goal:

\begin{equation}
W = W^*
\end{equation}

where $W^* = I_{out}^* / V_{in}$, and $I_{out}^*$ is referenced output current.

Let's assume that commutation parameters $d,s,\beta$ are at their steady-states, defined by controls \eqref{eq:d_inversion} and \eqref{eq:s_inversion} or corresponding special cases, discussed in the previous chapter. That means that coefficients $A$ and $B$ from \eqref{eq:FHA_coeffs} can be parametrized in terms of reference values $\sigma^*$ and $\delta^*$, along with $s_{add}$ (if this feature is used).

If the nonlinear inversion condition \eqref{eq:inversion_d_s} holds, then following trigonometric manipulations from section \ref{sec:inversion_condition}, it is possible to conclude that $B = A \tan \sigma^*$ and furthermore:

\begin{equation}
\sqrt{A^2 + B^2} = \frac{A}{\cos \sigma^*}
\end{equation}

The $A$ term is actually a function of reference values $\sigma^*, \delta^*, s_{add}$ and $G$ if the values $d,s,\beta$ are controlled to achieve the references:

\begin{equation}
\begin{gathered}
A = 4 \sin d(\sigma^*, \delta^*, s_{add}) + 4 G \sin(\sigma^* + \delta^* + s(\sigma^*, \delta^*, s_{add})) + 4 G \sin (\sigma^* + \delta^*) = A(\sigma^*, \delta^*, s_{add}, G)
\end{gathered}
\end{equation}

Based on this, the transconductance \eqref{eq:affine_model} can be separated into two parts $H$ and $Z$: one is parametrized by reference commutation parameters $\sigma^*, \delta^*, s_{add}$ and another one by switching frequency $\omega$.

\begin{equation}\label{eq:W_split}
W = \frac{I_{out}}{V_{in}} = \frac{n}{2 \pi^2} \frac{H(\sigma^*, \delta^*, s_{add}, G)}{Z(\omega)}  \\
\end{equation}

where 

\begin{equation}\label{eq:H}
H = A(\sigma^*, \delta^*, s_{add}, G) \cdot \frac{\cos(s(\sigma^*, \delta^*, s_{add}) + \delta^*) + \cos \delta^*}{\cos \sigma^*}
\end{equation}

Then the frequency control consists of calculation of $\omega$ which satisfies \eqref{eq:W_split} for a given referenced $W^*$. 
The \eqref{eq:W_split} can be solved with respect to $Z(\omega)$:

\begin{equation}\label{eq:Z}
Z = \frac{n}{2 \pi^2} \frac{H(\sigma^*, \delta^*, s_{add}, G)}{W^*}
\end{equation}

And in the same time $Z(\omega) = \omega L - 1/(\omega C)$, from here the required $\omega$ can be easilty obtained:

\begin{equation}\label{eq:omega}
\omega =  \frac{1}{2 L C} \left ( \sqrt{ C^2 Z^2 + 4 L C } + C Z \right )
\end{equation}

\subsection{Special case for fully driven converter}

The output transconductance with feed-forward maps $\cos \beta = G$ and $\delta  = 0$ can be calculated as after all simplifications using \eqref{eq:buck_beta} and $s = 0$: 

\begin{equation}
W = \frac{8 n}{\pi^2} \frac{1}{ Z(\omega) } \sqrt{ 1 - G^2 }
\end{equation}

Let's calculate frequency $\omega$ which is needed to achieve desired output $W^*$:

\begin{equation}
\begin{gathered}
Z = \frac{8 n \sqrt{ 1 - G^2 }} { \pi^2 W^* } \\
\omega =  \frac{1}{2 L C} \left ( \sqrt{ C^2 Z^2 + 4 L C } + C Z \right )
\end{gathered}
\end{equation}

The combined feed-forward control law (both for buck and boost modes) can be formulated as follows:

For given $W^*$ and $G$, and also pre-calculated buck to boost threshold $G^* = \cos (\sigma_{min})$:

1. First calculate the control inputs $\beta$ and $s$

\begin{equation}\label{eq:buck_boost_s_beta}
\begin{gathered}
\beta = \acos( \min \{ G, G^* \} ) \\
s = \acos \left ( \frac{2 G^*}{\max \{ G, G^* \}} - 1 \right )
\end{gathered}
\end{equation}

2. Then calculate desired tank impedance:

\begin{equation}
Z = 4 n \frac{\cos s + 1}{\pi^2 W^*} \sqrt{ 1 - G \cos(\beta + s) } 
\end{equation}

3. The switching frequency will be determined by equation:

\begin{equation}
\omega =  \frac{1}{2 L C} \left ( \sqrt{ C^2 Z^2 + 4 L C } + C Z \right )
\end{equation}

\subsection{Low power mode with $\omega = \omega_{max}$}\label{sec:freq_low_power}

The switching frequency in control \eqref{eq:omega} is unbounded, i.e. in order to get $W^* = 0$ frequency should be $\omega \to \infty$.
In real systems, the upper frequency is always constrained at $\omega_{max}$. Thus if we need output power even lesser than we can get at $\omega = \omega_{max}$, impedance can't be further increased and control of output power can be achieved only by commutation parameters $s$ and $\beta$ with fixed $\omega = \omega_{max}$. This mode we will call a low power mode.

Let's define maximum impedance that we can get:

\begin{equation}
Z_{max} = \omega_{max} L - \frac{1}{\omega_{max} C}
\end{equation}

The control law for low power can be formulated as follows: for a given $W^*$ find commutation parameters $d,s,\beta$ which are satisfying following conditions:

\begin{equation}\label{eq:control_problem_lp}
W^* = \frac{n}{2 \pi^2} \frac{H(\sigma^*, \delta^*, s_{add}, G)}{Z_{max}} 
\end{equation}

Let's consider an idea to limit output power by additional increase of $s$, which is equivalent to dimming the output down to zero power at $s = \pi$ when secondary side is completely shorted. Just to mention that an another possibility for output power regulation at constant frequency is to shift the secondary side respect to primary side by introducing $\delta > 0$.

For implementation of this idea the $s_{add}$ quantity was introduced in section \ref{sec:s_add} which resulted in inversion maps \eqref{eq:d_inversion} and \eqref{eq:s_inversion}. This mode of operation can be described as a partial rectification, because only a part of resonant tank current is being rectified by corresponding seconday side commutation, another part is being recycled and adds to reactive power. 

\begin{figure}
\centering
\includegraphics[width=1.0\textwidth]{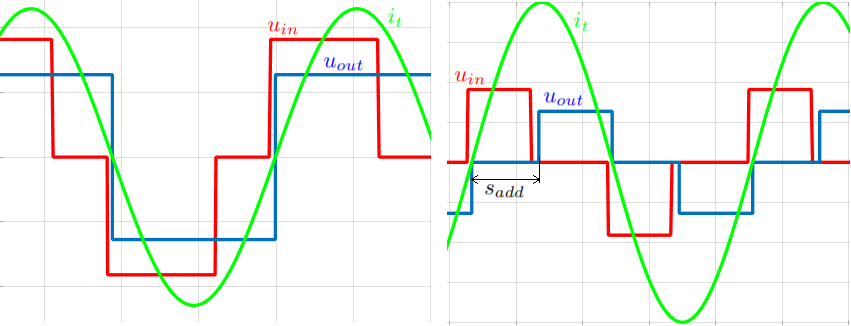}

\caption{Waveforms without (left) and with (right) an additional $s_{add}$ introduction for low power mode at $G = 0.7$}
\label{fig:LowPower_wave_examples}
\end{figure}

The examples of waveforms with and without of addition of $s_{add}$ are given in Figure \ref{fig:LowPower_wave_examples}. The operating point under consideration is defined by $G = 0.7$ (a buck mode), $\omega = 3\omega_0/2$, and $\delta^* = 0$, $\sigma^* = 0.1$. The output current calculated by \eqref{eq:affine_model} was 2 times lower when introducing $s_{add} = 3/2$ rad.

To characterize dependency of $W$ from $s$ with a fixed frequency, lets try to calculate $W$ by \eqref{eq:W_split} (or model equation \eqref{eq:affine_model}) for different values of $G$ and $s_{add}$. 
Let's denote $W_0$ the output power obtained for fixed maximal frequency $\omega_{max}$ without any additional secondary side shorting for low power, i.e. $s_{add} = 0$.

\begin{equation}
W_0 = \frac{n}{2 \pi^2} \frac{H(\sigma^*, \delta^*, 0, G)}{Z_{max}}
\end{equation}

\begin{figure}
\centering
\includegraphics[width=0.7\textwidth]{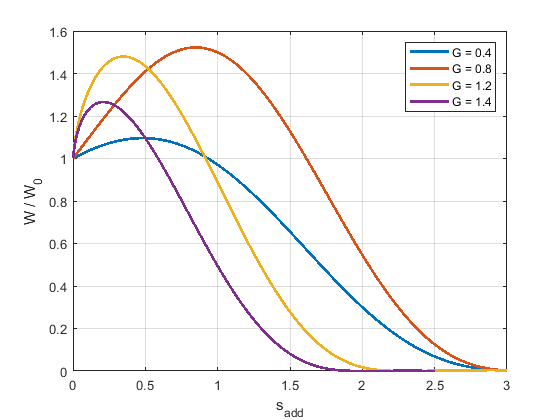}

\caption{Output power variation with $s_{add}$ for different $G$ and $\delta^* = 0$, $\sigma^* = 0.1$}
\label{fig:LowPower_curves}
\end{figure}

The output power variation for different values of G is demonstrated in Figure \ref{fig:LowPower_curves} as ratio of $W/W_0$. As one can see from the picture, a zero power operation $W = 0$ is achieved for $s = \pi$ (fully shorted secondary side) regardless of G. But an interesting property of the system is that $W$ doesn't decrease monotonically with increasing of $s_{add}$. That means that a discontinuous control should be applied for $s_{add}$ when transitioning from regular operation with $s_{add} = 0$ to a low power operation with $s_{add} > 0$ and $W < W_0$. 

When the low power control is requested, minimal $s_{add}^0 > 0$ needs to be applied which satisfies the condition $W(s_{add}^0) = W(0)$ or:

\begin{equation}\label{eq:s_add_0_condition}
H(\sigma^*, \delta^*, s_{add}^0, G) = H(\sigma^*, \delta^*, 0, G)
\end{equation}

\begin{figure}
\centering
\includegraphics[width=0.7\textwidth]{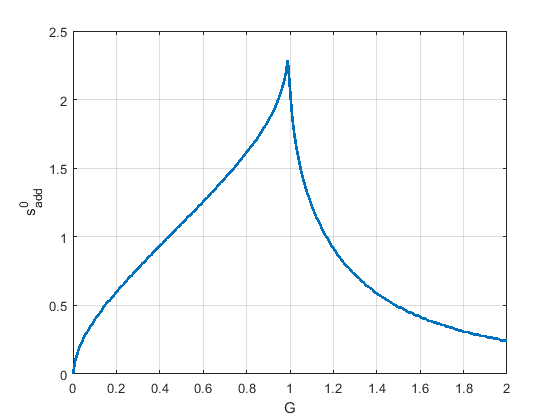}

\caption{Initial values of $s_{add}^0$ for monotonic decrease of $W$, the solution of \eqref{eq:s_add_0_condition}}
\label{fig:s_add_boundary}
\end{figure}

This $s_{add}^0$ is a value of $s_{add}$ after which the output power will decrease monotonically. The figure \ref{fig:s_add_boundary} illustrates the boundary of valid $s_{add}$ values, i.e. the solution of \eqref{eq:s_add_0_condition}.

\begin{figure}
\centering
\includegraphics[width=0.7\textwidth]{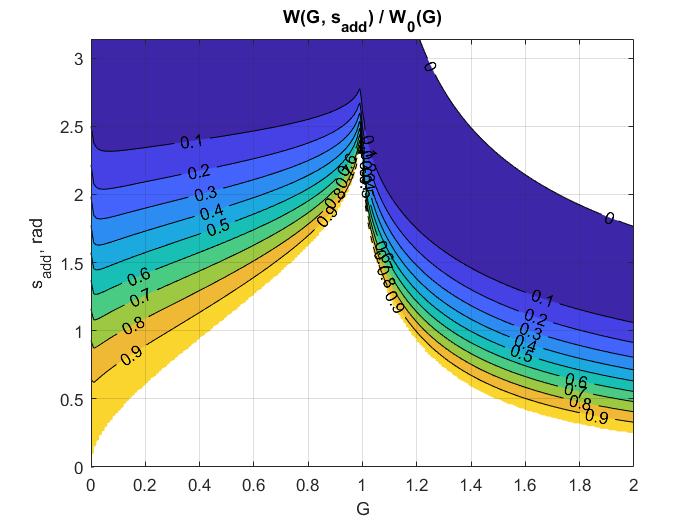}

\caption{Output power variation with $s_{add}$ for different $G$ and $\delta^* = 0$, $\sigma^* = 0.1$}
\label{fig:LowPower_boost}
\end{figure}

The full map of normilized $W$ to $W_0$ is given in the Figure \ref{fig:LowPower_boost}. The white space are areas of invalid $s_{add}$ values, where either $W > W_0$ or $W < 0$.

The inverse mapping algorithm of $W^{-1}$ for a given $\sigma^*$, $\delta^*$ and $W^*$ can be formulated as follows:

1. Calculate $\omega$ using \eqref{eq:Z} and \eqref{eq:omega} for $s_{add} = 0$ and determined commutation parameters $d,s,\beta$ obtained by \eqref{eq:d_inversion__s_add} and \eqref{eq:s_inversion__s_add}.

2. If $\omega > \omega_{max}$ by \eqref{eq:omega}, then use $\omega = \omega_{max}$ and iterate $s_{add}$ from $0$ to $\pi$ by recalculating the commutation parameters $d,s,\beta$ by \eqref{eq:d_inversion__s_add} and \eqref{eq:s_inversion__s_add} in order to approximately solve \eqref{eq:control_problem_lp} with respect to $s_{add}$.

\section{Controller architecture}

\subsection{Inner loop: commutation parameters control}

\subsubsection{Series nonlinear compensation}\label{sec:inner_loop_series}

In order to overcome uncertainties of the model \eqref{eq:FHA_coeffs}, additional external (linear) feedback controllers should be used in conjunction to decoupling and linearization as a solution of \eqref{eq:inv_problem}. 

\begin{figure}[H]
\input{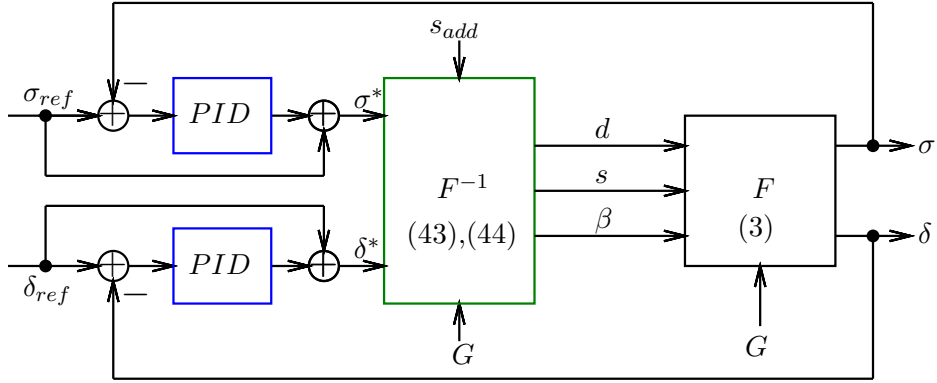}

\caption{Architecture of control system for $\sigma$ and $\delta$ with series nonlinear compensation.}
\label{fig:Inversion_reg}
\end{figure}

One approach would be to use a series nonlinear compensation, which consisting of regulating inputs $\sigma^*$ and $\delta^*$ of the nonlinear map $F^{-1}$ as it presented in Figure \ref{fig:Inversion_reg}. These two controllers could have PID form and regulate $\sigma$ and $\delta$ independently. The direct feedforward links added from the $\sigma$ and $\delta$ setpoints to the corresponding inputs of $F^{-1}$ map. Under an ideal model matching condition, the outputs of the PID regulator keep at zero. 

\subsubsection{Parallel nonlinear compensation}\label{sec:inner_loop_parallel}

It is well known [30], that the performance of series nonlinear compensation often susceptible to the plant uncertainties.
In order to overcome this issue, a parallel nonlinear compensation can be used. In this case the control action from feedback controllers is added to the outputs the feedback map $F^{-1}$. 

There are 3 outputs $d,s,\beta$ of the inversion map $F^{-1}$ in the particular system under the consideration, but two controllers for $\sigma$ and $\delta$.
In order to overcome this problem lets combine both $d$ and $s$ in a single variable $q$ by \eqref{eq:q}, using observation that if $s > 0$ then $d = \pi$ and if $s = 0$, then $d < \pi$ (assuming no low power mode, i.e. $s_{add} = 0$). In order to embed $s_{add}$ into the definition of $q$, the following modification of \eqref{eq:q} can be proposed:

\begin{equation}\label{eq:q__s_add}
\begin{pmatrix} d \\ s \end{pmatrix} = \begin{cases} \begin{pmatrix} q \\ s_{add} \end{pmatrix} \;,\; q \le \pi \\ \begin{pmatrix} \pi \\ q - \pi \end{pmatrix} \;,\; q > \pi \end{cases}
\end{equation}

Note that for $q > \pi$ in \eqref{eq:q__s_add}, the $s_{add}$ is added to $s$, so the equation consists $s_{add}$ only in the buck case $q < \pi$.

Let's consider two cases:

1. The buck mode $q < \pi$: $d = q$, $s = s_{add}$.

By substituting $d := q$ into \eqref{eq:inversion_d_s}, obtaining for $q \in [0,\pi]$: 

\begin{equation}\label{eq:q__buck}
q = \pm \acos (\cos \sigma^* - G \cos \delta^* - G \cos(\delta^* + s_{add})) + 2 \pi n + \sigma^*
\end{equation}

where $n$ is an integer number.

2. The boost mode $q > \pi$: $d = \pi$, $s = q - \pi = s_{min} + s_{add}$.

Substituting $s := s_{min} = q - \pi - s_{add}$ to \eqref{eq:inversion_d_s} and using the fact that $\cos(x - \pi) = \cos(\pi - x) = -\cos x$, then obtaining for $q \in [\pi,2\pi]$:

\begin{equation}\label{eq:q__boost_delta}
q = \pm \acos( \cos \delta^* - 2/G \cos \sigma^* ) + 2 \pi n - \delta^* + s_{add}
\end{equation}

where $n$ is an integer number.

Following this approach, the nonlinear inversion map becomes square, it has two inputs $\sigma^*$, $\delta^*$ and two outputs $q$, $\beta$.
One can observe that the $\sigma^*$ is a linear term of the $q$ for buck mode ($q < \pi$). So a feedback term which regulates $\sigma$ using additional linear controller can be added to $q$. 
A feedback term that regulates $\delta$ can be added to $\beta$ because it is linear for both $\delta$ and $\sigma$: $\beta = \delta + \sigma$.

The remaining part is $q$ in the boost mode \eqref{eq:q__boost_delta} which has $\delta^*$ as a linear term. This can be changed substituting $\delta^* = \beta - \sigma^*$.

\begin{equation}\label{eq:q__boost}
q = \pm \acos( \cos \delta^* - 2/G \cos \sigma^* ) + 2 \pi n + s_{add} - \beta + \sigma^*
\end{equation}

\begin{figure}[H]
\input{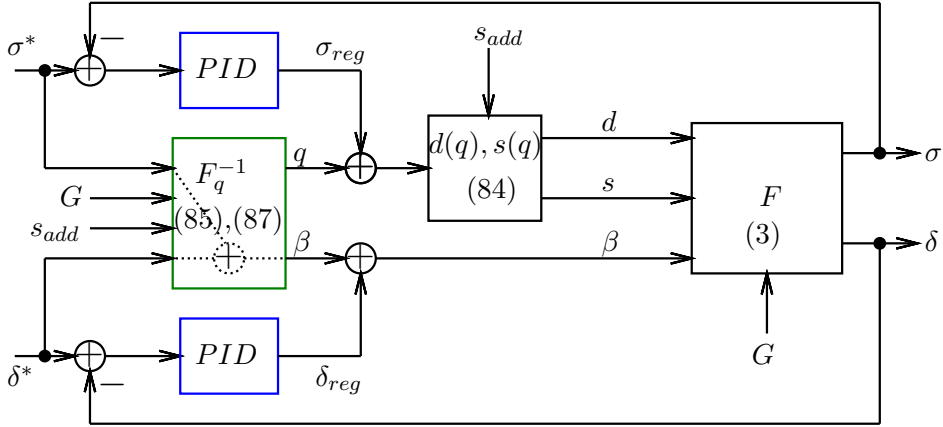}

\caption{Architecture of control system for $\sigma$ and $\delta$ with parallel nonlinear compensation.}
\label{fig:Inverstion_parallel}
\end{figure}

The resulted control architecture for $\sigma$ and $\delta$ is given in Figure \ref{fig:Inverstion_parallel}.

\subsection{Outer loop: output power control}

\subsubsection{Series nonlinear compensation}

The series nonlinear compensation essentially extends the architecture presented in section \ref{sec:inner_loop_series}. The mappings $W^{-1}$ and $F^{-1}$ are merged together since the calculation of frequency $\omega$ in accordance with the algorithm from the end of section \ref{sec:freq_low_power} requires actually calculated $d,s,\beta$ and iterating of $s_{add}$ which is an input of the mapping $F^{-1}$.

In a practical sense, the combined $W^{-1}$ and $F^{-1}$ map is a 3D look-up table.

\begin{figure}[H]
\input{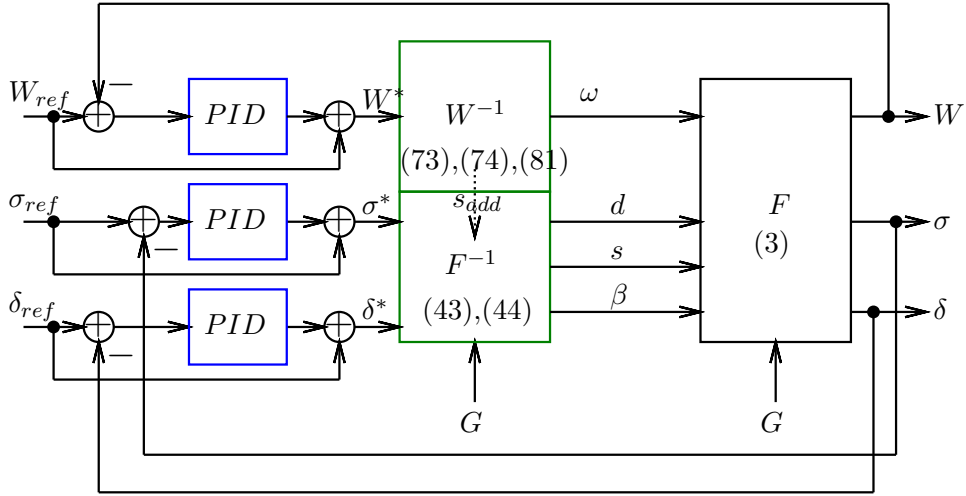}

\caption{Controller architecture with commutation parameters control $\sigma$ and $\delta$, and output power control $W$.}
\label{fig:Freq_Ctrl}
\end{figure}

\subsubsection{Overall architecture}\label{sec:outer_loop_parallel}

The final control architecture consists of parallel nonlinear compensation for commutation parameters $\sigma$ and $\delta$ as described in \ref{sec:inner_loop_parallel} with series nonlinear compensation of $\omega$ (and $s_{add}$) for regulation of $W$. As it evident from \eqref{eq:Z},\eqref{eq:omega}, the inverse map of $W$ is smooth and monotonic, so no need to derive parallel compensation scheme. This hybrid approach allows obtaining reasonably robust control, as it demonstrated in the next section with simulation results.

\begin{figure}[H]
\input{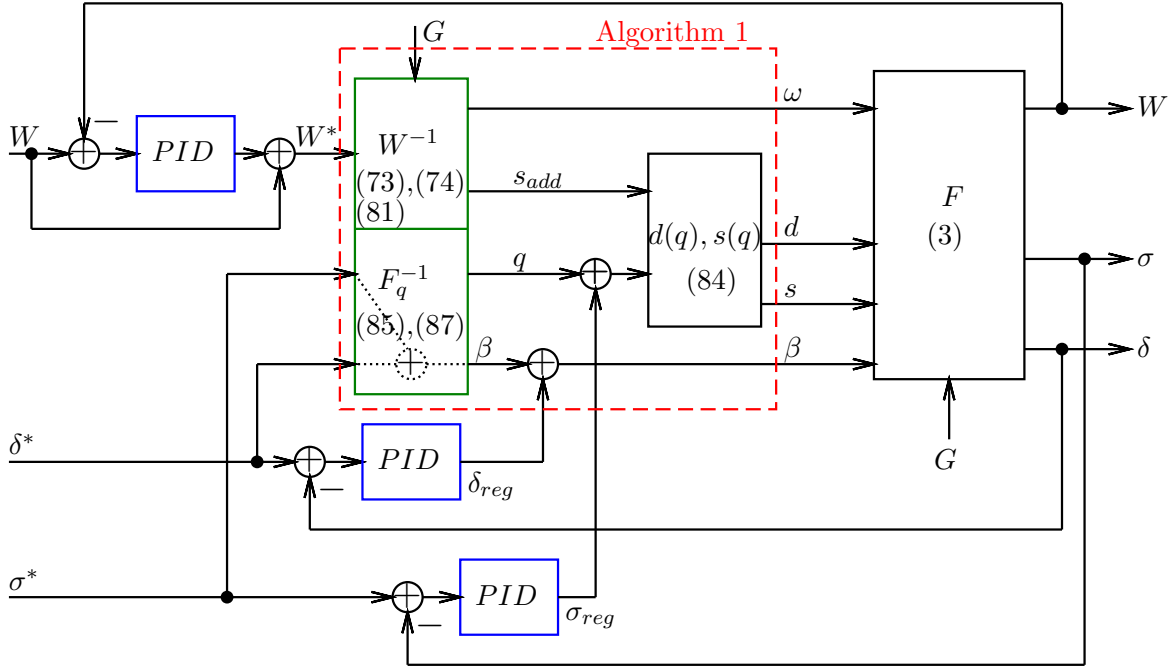}

\caption{Controller architecture with commutation parameters control $\sigma$ and $\delta$, and output power control $W$.}
\label{fig:Freq_Ctrl_parallel}
\end{figure}

The implementable algorithm of nonlinear inversion $F_q^{-1}$ and $W^{-1}$ can be formalized by the algorithm \ref{alg:inversion}.

\begin{algorithm}   
  \caption{Calculation of inversion maps $F_q^{-1}$ and $W^{-1}$.}
  \label{alg:inversion}
  
\begin{algorithmic}

\ForAll {$s_{add} \in [0, \pi]$}
    \State $x := \cos \sigma^* - G \cos \delta^* - G \cos(\delta^* + s_{add})$ \Comment{trying to evaluate \eqref{eq:q__buck}}
    \If {$|x| \le 1$} \Comment{if this is a buck mode?}
        \State $q := \acos x + \sigma^*$ \Comment{calculate $q$ for buck mode}
    \Else \Comment{else it is a boost mode}
        \State $q := 2 \pi - \acos( \cos \delta^* - 2/G \cos \sigma^* ) - \delta^* + s_{add}$ \Comment{calculate $q$ for boost mode by \eqref{eq:q__boost}}
    \EndIf

    \State $\beta := \sigma^* + \delta^*$  \Comment{calculate $\beta$ by \eqref{eq:beta}}
    \State $\beta := \beta + \delta_{reg}$ \Comment{add external PID controller action for $\delta$}
    \State $q := q + \sigma_{reg}$  \Comment{add external PID controller action for $\sigma$}

    \If {$q \le \pi$} \Comment{determining $d,s$ depending if this is buck or boost}
        \State $d := q$  \Comment{it is a buck mode, so $d \le \pi$}
        \State $s := s_{add}$ \Comment{in case if $s_{add}$ requested for low power}
    \Else 
        \State $d := \pi$  \Comment{it is a boost mode, so $d = \pi$}
        \State $s := q - \pi$  \Comment{$s_{add}$ is contained in $s$}
    \EndIf

    \State $A := 4 \sin d + 4 G \sin(\beta + s) + 4 G \sin \beta$   \Comment{by \eqref{eq:FHA_coeffs}}
    \State $H := A \cdot \dfrac{\cos(s + \delta^*) + \cos \delta^*}{\cos \sigma^*}$    \Comment{by \eqref{eq:H}}
    \State $Z := \dfrac{n}{2 \pi^2} \dfrac{H}{W^*}$    \Comment{by \eqref{eq:Z}}
    \State $\omega := \dfrac{1}{2 L C} \left ( \sqrt{ C^2 Z^2 + 4 L C } + C Z \right )$    \Comment{by \eqref{eq:omega}}

    \If {$\omega \le \omega_{\max}$}
        \Return \Comment{switching parameters are calculated}
    \Else 
        \State $s_{add} := s_{add} + \Delta s_{add}$  \Comment{increase $s_{add}$ by a small step $\Delta s_{add}$}
    \EndIf
\EndFor

\end{algorithmic}

\end{algorithm}

\section{Simulation}

A MATLAB Simulink model was developed in order to prototype the control strategies proposed.
A battery charger application is used as case study example.

A simple model of 30Ah pack is used as a load of the resonant converter.
The battery model consists of an integrator for coulomb counting and a map which maps Ah to battery voltage.

\begin{figure}
\centering
\includegraphics[width=0.8\textwidth]{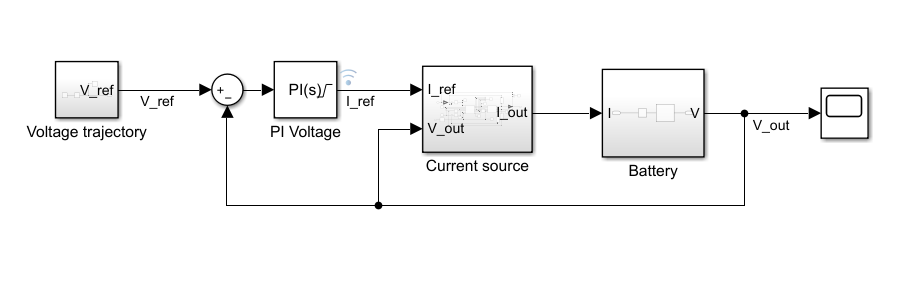}

\caption{Top view of EV charger Simulink model}
\label{fig:Sim__ModelTop}
\end{figure}

The charger is a power supply that consists of an outer voltage loop which regulates the current and inner current loop with input limitation.
The model is presented in Figure \ref{fig:Sim__ModelTop}. The submodel block \textit{Current source} is a model of closed loop resonant LC DC/DC converter that implements the control approach developed in this paper.

\begin{figure}
\centering
\includegraphics[width=1.0\textwidth]{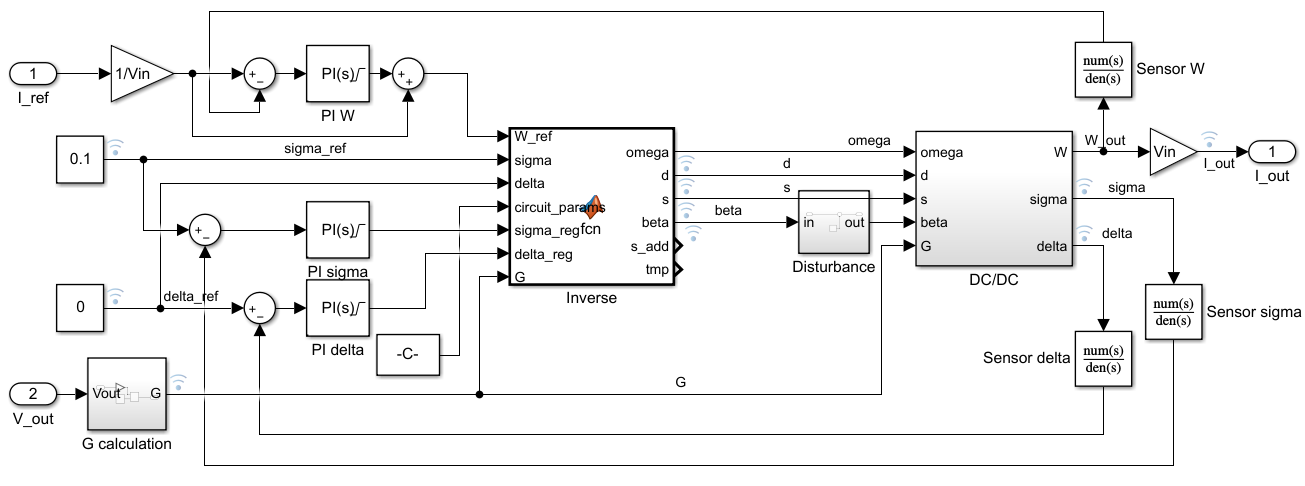}

\caption{The Simulink model for current source, the LC resonant DC/DC converter}
\label{fig:Sim__ModelCurr}
\end{figure}

The internals of \textit{Current source} block is shown in Figure \ref{fig:Sim__ModelCurr}

The block \textit{Inverse} implements the algorithm \ref{alg:inversion}. The block \textit{DC/DC} models steady-state model \eqref{eq:FHA_coeffs} of LC converter and reused from [23]. The sensor dynamics for measurements of $W$, $\sigma$ and $\delta$ are modeled by first order transfer functions.

The parameters of the resonant tank were given as follows: $C = 47$ nF, $L = 80$ uH, which gives the resonant frequency of about 82 kHz. The maximum frequency is set to 165 kHz.

The transformer turns ration is chosen so that the point of $G = 1$ is being passed roughly in the middle of the charge. The charging profile consists of constant current (CC) mode with 25 A setpoint followed by a constant voltage (CV) mode with 400 V setpoint. The voltage varies during the charge from 240 V to 400 V. The input voltage of the converter is fixed at 600 V. 

In order to demonstrate robustness of the proposed control approach two model uncertainties were introduced:

1. Constant offset is applied to the $\beta$ switching parameter at the plant's (converter's) side: $\beta := \beta - 0.1$.

2. The series inductance of the resonant tank is increased by 5\%.

\begin{figure}
\centering
\includegraphics[width=0.7\textwidth]{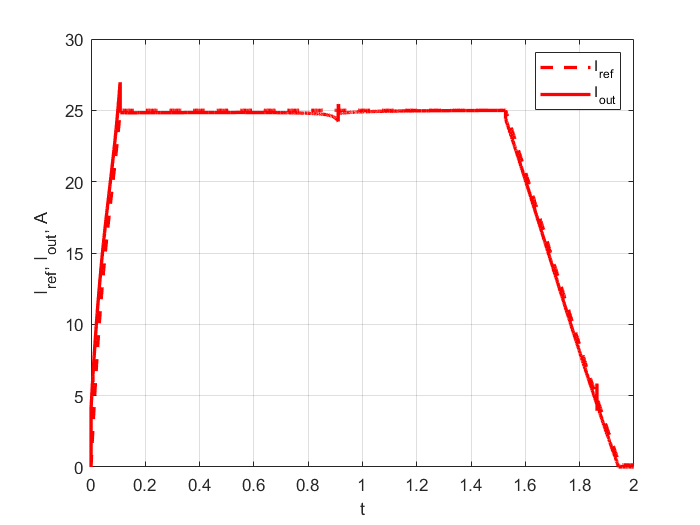}

\caption{Output and referenced current}
\label{fig:Sim__I}
\end{figure}

\begin{figure}
\centering
\includegraphics[width=0.7\textwidth]{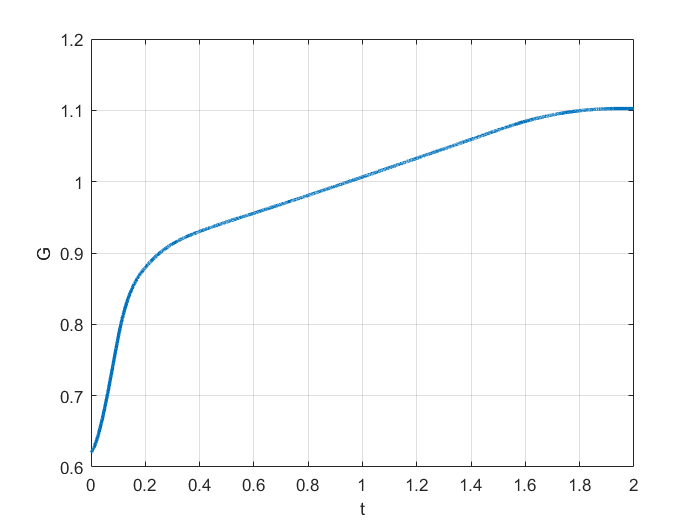}

\caption{Variation of the voltage ratio $G$ during the charge}
\label{fig:Sim__G}
\end{figure}

\begin{figure}
\centering
\includegraphics[width=0.7\textwidth]{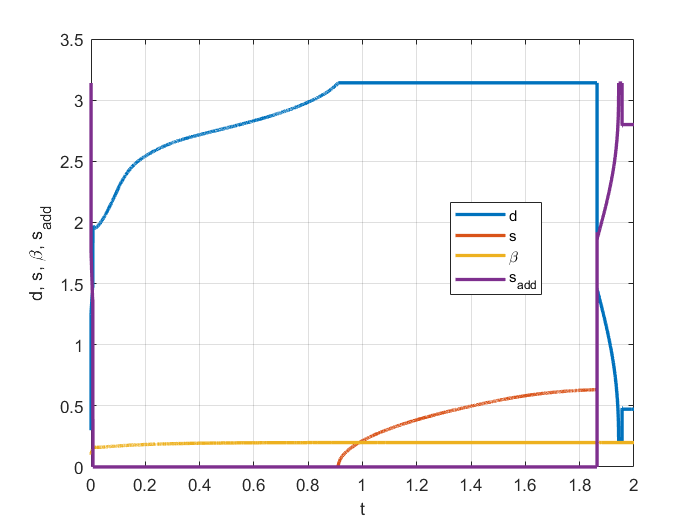}

\caption{The control outputs for commutation PWM parameters}
\label{fig:Sim__d_s_beta}
\end{figure}

\begin{figure}
\centering
\includegraphics[width=0.7\textwidth]{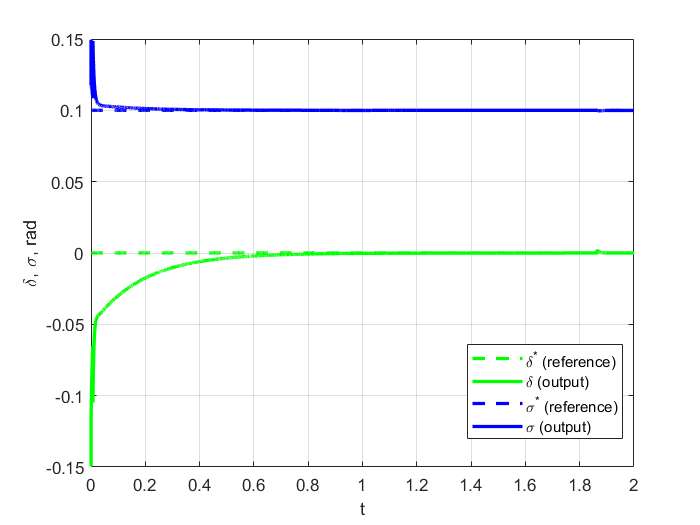}

\caption{Controllable outputs of waveform alignment quantities and their reference values}
\label{fig:Sim__sigma_delta}
\end{figure}

The results of simulation presented in Figures \ref{fig:Sim__I}, \ref{fig:Sim__G}, \ref{fig:Sim__d_s_beta} and \ref{fig:Sim__sigma_delta}. The time axis has arbitrary scale and the dynamics is subject to PI tuning.

From the simulation it is evident that the controllable outputs are following their reference values while the converter passes through lower power buck mode, regular buck mode, then boost mode and finally ends up with low power boost mode. The plant uncertainty introduced are correctable by external PI controllers in accordance to the architecture proposed in the section \ref{sec:outer_loop_parallel}.

\section{Conclusion}

In this paper a control approach that consists of direct nonlinear inversion of the steady-state model equations \eqref{eq:FHA_coeffs} under various operating conditions.

Namely, the following main results are obtained:

1. The optimality of operating mode is characterized using angular quantities \eqref{eq:sigma_delta_intro} that measure alignment of the primary and secondary switching event with respect to the tank current. The main novelty of the approach presented, that these phase displacements are being measured and controlled using feedback controllers, which maintains the efficiency of the converter in a wide operating range and in presence of uncertainties.

2. The nonlinear control problem is formulated \eqref{eq:inv_problem} and solved \eqref{eq:d_inversion__s_add}, \eqref{eq:s_inversion__s_add} to actively control the waveform alignment quantities. As a result, a ZVS commutation can be maintained across a wide range of operating conditions.

3. An indirect synchronous rectification approach is a byproduct of the active control of the alignment of the secondary side voltage waveform with the resonant tank current. This approach increases the converter efficiency by a compensation of gate and sensing circuit delays, and is more robust to a noise of the zero current detection circuit.

4. The nonlinear inversion is square, i.e. has the same number of inputs and outputs. It is achieved by combining duty cycles of primary and secondary sides into a single variable \eqref{eq:q__s_add}.

5. The output current regulation problem \eqref{eq:W_split} is formulated in terms of transconductance, which normalizes the converter operation by the input voltage.

6. For the low power operation, an additional duty cycle modulation of the secondary side under constant frequency is proposed \eqref{eq:control_problem_lp} and corresponding control strategy is derived.

7. A parallel nonlinear compensation which consists of adding control actions of external linear regulators to the manipulated plant inputs is developed in order to achieve robustness of the control implementation.

The future work will be devoted to a cycle-to-cycle dynamics of the converter and corresponding nonlinear control.

\end{document}